\documentclass{IEEEtran}
\usepackage{algorithm}
\usepackage{algpseudocode}
\usepackage[utf8]{inputenc}

\usepackage{amsmath,amssymb,epsfig,psfrag,cite,subfigure}
\include{macros}
\usepackage{graphicx}
\usepackage{epstopdf}
\usepackage{balance}
\usepackage{acronym}

\usepackage{bm}

\usepackage{geometry}
\usepackage{tcolorbox}

\usepackage{accents}
\allowdisplaybreaks
\usepackage{soul}

\usepackage{amsthm}

\usepackage{nicefrac}

\usepackage{lipsum,multicol}

\usepackage{algorithm}
\usepackage{algpseudocode}
\algdef{SE}[SUBALG]{Indent}{EndIndent}{}{\algorithmicend\ }%
\algtext*{Indent}
\algtext*{EndIndent}

\usepackage{enumitem}
\usepackage{mwe}    
\usepackage{epsfig,psfrag}
\usepackage{subfigure}
\usepackage{color}
\usepackage{url}
\usepackage{mathtools,xparse}

\usepackage{gensymb}

\usepackage[multiple]{footmisc}

\usepackage{xr}
\makeatletter
\newcommand*{\addFileDependency}[1]{% argument=file name and extension
  \typeout{(#1)}
  \@addtofilelist{#1}
  \IfFileExists{#1}{}{\typeout{No file #1.}}
}
\makeatother

\newcommand*{\myexternaldocument}[1]{%
    \externaldocument{#1}%
    \addFileDependency{#1.tex}%
    \addFileDependency{#1.aux}%
}

\myexternaldocument{supp}

\usepackage{xcolor}

\usepackage[T1]{fontenc}
\usepackage[utf8]{inputenc}
\usepackage{pgfplots}
\usepackage{grffile}
\pgfplotsset{compat=newest}
\usetikzlibrary{plotmarks}
\usepgfplotslibrary{patchplots}
\usepackage{amsmath}
\DeclareMathOperator*{\argmax}{arg\,max}
\DeclareMathOperator*{\argmin}{arg\,min}

\newcommand{\mr}[1]{\mathrm{#1}}
\DeclareMathOperator{\atant}{\mr{atan2}}

\newcommand\abs[1]{\big|#1\big|}

\newcommand\norm[1]{\lVert #1\rVert}

\newcommand{\pp}{\boldsymbol{p}}
\newcommand{\vv}{\boldsymbol{v}}
\newcommand{\hh}{\boldsymbol{h}}
\newcommand{\ww}{\boldsymbol{w}}
\newcommand{\qq}{\boldsymbol{q}}

\newcommand{\zz}{\boldsymbol{z}}

\newcommand{\rr}{\boldsymbol{r}}

\newcommand{\bb}{\boldsymbol{b}}

\newcommand{\bw}{\boldsymbol{w}}

\newcommand{\bi}{\boldsymbol{i}}
\newcommand{\uu}{\boldsymbol{u}}
\newcommand{\bz}{\boldsymbol{z}}

\newcommand{\bF}{\mathbf{F}}

\newcommand{\bc}{\mathbf{c}}
\newcommand{\bN}{\mathbf{N}}

\newcommand{\bH}{\mathbf{H}}
\newcommand{\bI}{\mathbf{I}}

\newcommand{\bn}{\mathbf{n}}

\newcommand{\chibch}{\Chib_{\rm{ch}}}

\newcommand{\thn}[1]{ {#1^{\rm{th} } } }

\newcommand{\bA}{\boldsymbol{A}}
\newcommand{\bJ}{\boldsymbol{J}}

\newcommand{\bY}{\mathbf{Y}}
\newcommand{\bZ}{\boldsymbol{Z}}
\newcommand{\bR}{\boldsymbol{R}}

\newcommand{\bq}{\boldsymbol{q}}

\newcommand{\bD}{\mathbf{D}}

\newcommand{\by}{\mathbf{y}}

\newcommand{\aab}{\boldsymbol{a}}

\newcommand{\Chib}{\bm{\chi}}
\newcommand{\alphareBS}{\alpha_{{\rm{R}},\BS}}
\newcommand{\alphaimBS}{\alpha_{{\rm{I}},\BS}}
\newcommand{\alphareSAT}{\alpha_{{\rm{R}},\SAT}}
\newcommand{\alphaimSAT}{\alpha_{{\rm{I}},\SAT}}
\newcommand{\Ent}{i}

\newcommand{\ppbs}{\pp_{\BS}}

\newcommand{\ppsat}{\pp_{\SAT}}
\newcommand{\ppsato}{\pp_\SAT}

\newcommand{\UE}{\text{u}}
\newcommand{\SAT}{\text{s}}
\newcommand{\BS}{\text{b}}

\newcommand{\LEO}{\text{LEO}}
\newcommand{\E}{\text{E}}

\newcommand{\res}{\text{res}}

\newcommand{\bs}{\text{sb}} % weird issue with the next line
\newcommand{\ch}{\text{ch}}
\newcommand{\pos}{\text{pos}}
\newcommand{\el}{\text{el}}

\newcommand{\Ts}{T_s}
\newcommand{\Tcp}{T_\text{cp}}

\newcommand{\complexset}[2]{ \mathbb{C}^{#1 \times #2}  }
\newcommand{\complexsetone}[1]{ \mathbb{C}^{#1}  }

\newcommand{\realsetone}[1]{ \mathbb{R}^{#1}  }

\newcommand{\her}{\mathsf{H}}
\newcommand{\trp}{\mathsf{T}}

\newcommand{\conj}{*}

\newcommand{\thetab}{\bm{\theta}}

\newcommand{\thetaaz}{\theta_{\rm{az}}}
\newcommand{\thetael}{\theta_{\rm{el}}}

\makeatletter \renewcommand\d[1]{\ensuremath{%
		\;\mathrm{d}#1\@ifnextchar\d{\!}{}}}
\makeatother

\makeatletter
\newcommand*\rel@kern[1]{\kern#1\dimexpr\macc@kerna}
\newcommand*\widebar[1]{%
  \begingroup
  \def\mathaccent##1##2{%
    \rel@kern{0.8}%
    \overline{\rel@kern{-0.8}\macc@nucleus\rel@kern{0.2}}%
    \rel@kern{-0.2}%
  }%
  \macc@depth\@ne
  \let\math@bgroup\@empty \let\math@egroup\macc@set@skewchar
  \mathsurround\z@ \frozen@everymath{\mathgroup\macc@group\relax}%
  \macc@set@skewchar\relax
  \let\mathaccentV\macc@nested@a
  \macc@nested@a\relax111{#1}%
  \endgroup
}
\makeatother

\newtheorem{remark}{Remark}

\acrodef{RIS}{reconfigurable intelligent surface}
\acrodef{SNR}{signal-to-noise ratio}
\acrodef{ISAC}{integrated sensing and communication}
\acrodef{ISLAC}{integrated sensing, localization, and communication}
\acrodef{LoS}{line-of-sight}
\acrodef{NLoS}{non-line-of-sight}
\acrodef{AoA}{angle-of-arrival}
\acrodef{ToA}{time-of-arrival}
\acrodef{AoD}{angle-of-departure}
\acrodef{UE}{user equipment}
\acrodef{NF}{near-field}
\acrodef{BS}{base station}
\acrodef{MCRB}{misspecified Cram\'{e}r-Rao bound}
\acrodef{CRB}{Cram\'{e}r-Rao bound}
\acrodef{LB}{lower bound}
\acrodef{ML}{maximum-likelihood}
\acrodef{MML}{mismatched maximum-likelihood}
\acrodef{DL}{downlink}
\acrodef{UL}{uplink}
\acrodef{MIMO}{multiple-input multiple-output}
\acrodef{MISO}{multiple-input single-output}
\acrodef{SISO}{single-input single-output}
\acrodef{SIP}{shift invariance property}
\acrodef{FIM}{Fisher information matrix}
\acrodef{RMSE}{root mean-squared error}
\acrodef{AWGN}{additive white Gaussian noise}
\acrodef{ADMM}{alternating direction method of multipliers}
\acrodef{LS}{least-squares}
\acrodef{SOC}{second-order cone}
\acrodef{CFO}{carrier frequency offset}
\acrodef{GLRT}{generalized likelihood ratio test}
\acrodef{NTN} {non-terrestrial network}
\acrodef{TN} {terrestrial network}
\acrodef{LEO}{low Earth orbit}
\acrodef{GNSS}{global navigation satellite system}
\acrodef{OFDM}{orthogonal frequency division multiplexing}
\acrodef{UPA}{uniform planar array}
\acrodef{ICI}{inter-carrier interference}
\acrodef{HAPS}{high altitude platform station}
\acrodef{UAV}{unmanned aerial vehicle}
\acrodef{FSPL}{free space path-loss}
\acrodef{GDoP}{geometric dilution of precision}
\acrodef{FDoA}{frequency difference of arrival}
\acrodef{AD}{autonomous driving}
\acrodef{ADAS}{advanced driver-assistance systems}
\acrodef{RRC}{root-raised cosine}
\acrodef{IMU}{inertial measurement unit}
\usepackage{balance}

\allowdisplaybreaks
\begin{document}
\bstctlcite{IEEEexample:BSTcontrol}

%%%%%%%%%%%%%%%%%% title page information %%%%%%%%%%%%%%%%%%
% \title{Coordinated 5G LEO-5G Cellular Positioning for Improved Safety Functions in Heavy Vehicles}
\title{Integrated Cellular and LEO-based Positioning and Synchronization under User Mobility}
\author{Yasaman Ettefagh,~\IEEEmembership{Student Member,~IEEE}, Sharief Saleh,~\IEEEmembership{Member,~IEEE}, Musa Furkan Keskin,~\IEEEmembership{Member,~IEEE}, Hui Chen,~\IEEEmembership{Member,~IEEE}, Gonzalo Seco-Granados,~\IEEEmembership{Fellow,~IEEE}, and Henk Wymeersch,~\IEEEmembership{Fellow,~IEEE}
\thanks{This work is supported by Vinnova FFI project 2023-02603 and by the Swedish Research Council (VR) under Grants 2022-03007 and 2024-04390, and in part by Grant PID2023-152820OB-I00 funded by MICIU/AEI/10.13039/501100011033 and by ERDF/EU, and the AGAUR - ICREA Academia Programme.

Yasaman Ettefagh is with the Department of Electrical Engineering, Chalmers University of Technology, 41296 Gothenburg, Sweden and also Volvo Technology AB, Gothenburg, Sweden (e-mail:ettefagh@chalmers.se; yasaman.ettefagh@volvo.com). 

Sharief Saleh, Musa Furkan Keskin, Hui Chen and Henk Wymeersch are with the Department of Electrical Engineering, Chalmers University of Technology, 41296 Gothenburg, Sweden (emails: sharief@chalmers.se; furkan@chalmers.se; hui.chen@chalmers.se; henkw@chalmers.se). 

Gonzalo Seco-Granados is with the Department of Telecommunications and Systems Engineering, Universitat Autonòma de Barcelona, 08193 Bellaterra,
Spain (e-mail: gonzalo.seco@uab.cat).}}

\maketitle

%%%%%%%%%%%%%%%%%% abstract %%%%%%%%%%%%%%%%%%
\begin{abstract}
This paper investigates the localization, synchronization, and speed estimation of a mobile \ac{UE} leveraging integrated terrestrial and \acp{NTN}, in particular \ac{LEO} satellites. We focus on a minimal setup in which the \ac{UE} received signal from only one \ac{BS} and one \ac{LEO} satellite. %The \ac{BS} is equipped with a \ac{UPA}, while the satellite and \ac{UE} each utilize a single antenna. 
We derive a generic signal model accounting for mobility, clock and frequency offsets, based on which a hierarchy of simplified models are proposed and organized by computational complexity. Estimation algorithms are developed for each model to facilitate efficient and accurate parameter recovery. Rigorous simulations validate the effectiveness of the proposed models, demonstrating their suitability across diverse scenarios. The findings highlight how the trade-off between complexity and performance can be optimized for varying deployment environments and application requirements, offering valuable insights for 6G positioning and synchronization systems under user mobility.
\end{abstract}


% \tableofcontents
\acresetall 
\vspace{-5mm}
\section{Introduction}
% \subsection{P1 : sth about 6G and NTN - can delegate}
% \subsection{P2: importance of 6G positioning - can delegate}
% \subsection{P3: limitation of TN and NTN positioning separately - Sharief mag}
% \subsection{P4: Integration of TN and NTN - me}
% \subsection{P5: In this paper ... - me}

% P1:
\IEEEPARstart{N}{on}-terrestrial networks (NTNs) are becoming a critical component in the evolution of wireless communication systems, particularly in the transition from 5G to 6G. By integrating satellite systems, \ac{HAPS}, and airborne networks, NTNs aim to provide ubiquitous and seamless global coverage, addressing connectivity gaps in under-served and remote regions \cite{araniti2021toward,jiang2021road}. Beyond communication, NTNs have the potential to transform localization services by enabling global, high-precision positioning that is essential for a variety of emerging 6G applications \cite{azari2022evolution, dureppagari2023ntn}. Among various NTN technologies, \ac{LEO} satellites stand out due to their low latency, high capacity, and scalability through large constellations \cite{azari2022evolution,guidotti2022path}. These capabilities make LEO satellites 
 effective for critical localization applications such as \ac{AD} and \ac{ADAS}, where accuracy, reliability, and global availability are essential \cite{jiang2021road,guidotti2022path}. By complementing terrestrial systems, NTNs are shaping the future of integrated localization and communication, enabling safe and efficient operation of autonomous vehicles, industrial automation, and other next-generation services \cite{araniti2021toward,azari2022evolution}.
%\IEEEPARstart{N}{on}-terresterial networks (NTNs) are emerging as a transformative component of wireless communication systems, particularly in the evolution from 5G to 6G. By integrating satellite systems, high-altitude platforms (HAPs), and airborne networks, NTNs aim to provide seamless global coverage, effectively addressing connectivity gaps in remote and underserved areas. Among these technologies, \ac{LEO} satellites stand out due to their low latency, high capacity, and scalability through large constellations, making them a key enabler for achieving the ambitious goals of 6G networks. \textit{Also add} .

% P2:
Positioning is increasingly central to the vision of 6G, enabling a wide range of advanced applications \cite{behravan2022positioning}. As part of this evolution, the design of integrated positioning systems that leverage the complementary strengths of terrestrial and non-terrestrial infrastructures has emerged as a key focus in realizing the next-generation communication and sensing ecosystem \cite{jiang2021road, saleh2025integrated6gtnntn}. Recent research has explored cellular localization frameworks that emphasize integrity, fault tolerance, and robust error bounding for safety-critical applications \cite{whiton2022cellular}. In parallel, hybrid positioning approaches that integrate terrestrial 5G and non-terrestrial systems, such as GNSS, have demonstrated promising sub-meter accuracy in field trials \cite{del2023preliminary}. These applications demand high reliability and accuracy to ensure safe operation under diverse and challenging conditions, including urban environments with severe multipath interference and remote areas with limited terrestrial coverage.
\begin{figure}
    \centering
    \includegraphics[trim=0pt 0pt 10pt 30pt,clip,width=1\linewidth]{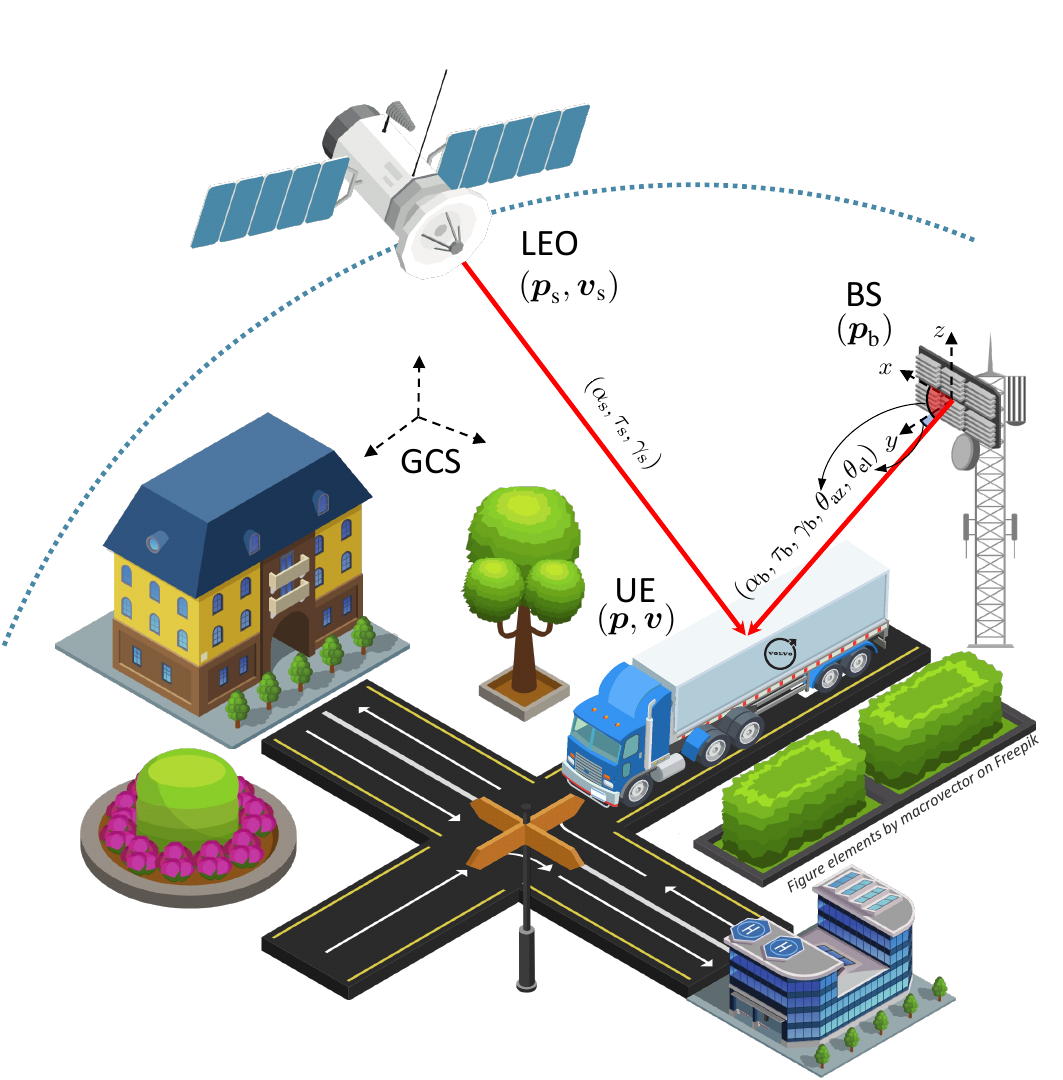}
    \caption{TN-NTN integrated setup: the mobile UE is localized from coordinated downlink signals from the fixed BS and the mobile LEO satellite. }
    \label{fig_NTN_setup}
\end{figure}

% P3: 
Terrestrial cellular networks have been widely utilized for positioning services, but they face several critical limitations that impact accuracy and reliability \cite{dwivedi2021positioning}. %A major challenge is multipath propagation and \ac{NLoS} conditions, especially in urban and indoor environments, where signal reflection and diffraction cause significant errors in distance and angle estimations \cite{xie2024positioning}. 
%The limited bandwidth of cellular networks further reduces the ability to resolve multipath components, amplifying errors in time delay and \ac{AoA} measurements. %Additionally, the geometric distribution of \acp{BS} plays a crucial role in positioning accuracy. Sparse or irregular \ac{BS} placements result in poor geometry, negatively affecting the \ac{DoP} and making precise localization difficult in rural or less densely populated areas.
%
%Another challenge is inter-cell interference, which arises from the frequency reuse schemes in cellular systems. At the edges of cells, interference from neighboring \acp{BS} often masks weaker signals, limiting the hearability of multiple \acp{BS} and degrading the performance of trilateration-based methods. 
Precise synchronization between \acp{BS} and mobile devices is required for accurate positioning, yet many cellular networks lack the stringent synchronization mechanisms needed for high-precision localization \cite{li2017analysis}. Another challenge is the limited coverage, as most positioning solutions require connection to at least four \acp{BS}, placing high demands on the infrastructure, compared to communication \cite{del2017survey}. 
%
%These challenges are particularly pronounced in methods like Enhanced Cell-ID and OTDoA, which rely on sufficient signal quality and neighbor BS detectability.
% Finally, the performance of cellular positioning in indoor scenarios remains limited due to high signal attenuation, rich multipath environments, and the lack of synchronized small-cell deployments \cite{del2017survey}.
%
%Terrestrial cellular networks have been extensively utilized for positioning services; however, they face inherent limitations that affect accuracy and reliability. A significant challenge is the reliance on signal measurements, such as \ac{ToA} and \ac{AoA}, which can be compromised by multipath propagation and \ac{NLoS} conditions prevalent in urban and indoor environments. These factors lead to signal reflections and scattering, causing errors in distance and angle estimations. Additionally, the geometric distribution of \acp{BS} influences positioning accuracy; sparse or irregular BS placement can degrade performance, especially in rural or less densely populated areas. Moreover, the limited bandwidth allocated for positioning in cellular networks constrains the achievable resolution, further impacting accuracy. These challenges necessitate the exploration of alternative or supplementary positioning systems to meet the stringent requirements of emerging applications \cite{del2017survey}. 
%
On the other hand, \ac{LEO} satellite systems offer global coverage and have been proposed as a solution to enhance positioning services, considering both opportunistic form \cite{neinavaie2021acquisition} and as part of 6G \cite{dureppagari2023ntn,dureppagari2023ntn-arxiv}. However, they also encounter specific limitations. The high mobility of \ac{LEO} satellites leads to rapidly changing satellite geometries, causing high Doppler shifts \cite{ali1998doppler}. % and frequent handovers between satellite beams, which complicate the maintenance of continuous and accurate positioning. 
% Early NTN deployments typically use sparse constellations optimized for coverage rather than localization, resulting in poor geometry for positioning and high \ac{GDoP}. Additionally, \ac{LEO}-based positioning relies heavily on signal measurements, such as time of arrival \ac{ToA} and \ac{FDoA}, which are hampered by extended propagation delays and delay spreads between satellites.%, exceeding the tolerances of current frameworks. Implementing FDOA also requires redesigning positioning reference signals (PRS) to provide sufficient Doppler resolution, a feature not yet supported in most NTN systems.
Furthermore, the beam patterns of LEO satellites, designed for communication purposes, often have narrow beam-widths for maximizing spectral efficiency, which conflicts with the broader, overlapping beams needed for accurate positioning. Interference management between overlapping beams in dense LEO constellations further complicates the system design \cite{dureppagari2023ntn}.

Integrating terrestrial and NTN %for positioning 
has attracted significant research interest due to the complementary strengths of these systems, both for communication  \cite{harounabadi2023toward,geraci2022integrating,charbit2021satellite} and for localization \cite{sallouha2024ground,gonzalez20235g,gonzalez2024interference,liang2024toward,jin2024fusion, saleh2025integrated6gtnntn}. Cellular networks, with their dense infrastructure and advanced signal processing capabilities, provide reliable positioning in urban areas, whereas NTNs, particularly \ac{LEO} satellites, offer global coverage and resilience in rural or \ac{GNSS}-denied environments. Recent studies have demonstrated the potential of 5G positioning reference signals to support integrated terrestrial-NTN scenarios, addressing challenges like Doppler shifts and interference in \ac{LEO}-based networks \cite{gonzalez20235g,gonzalez2024interference}.
%
%Furthermore, the vision for 6G includes a seamless convergence of communication, sensing, and localization functionalities to support advanced applications like autonomous systems and immersive extended reality \cite{saad2019vision}. 
Fusion-based approaches, such as combining pseudorange and \ac{AoA} measurements, have shown promise in enhancing positioning accuracy under low observability \cite{jin2024fusion}. This integrated framework can bridge the performance gaps of individual systems, providing robust, precise positioning for emerging 6G applications.
\begin{comment}
    
\begin{table}[]
    \centering
    \begin{tabular}{c|c}
         &  \\
         & 
    \end{tabular}
    \caption{Overview of the related works}
    \label{tab_literature}
\end{table}
\end{comment}

% In, \cite{saad2019vision} integration of terrestrial, airborne and satellite networks is considered as one of the technology-enablers in 6G. 

%\textit{Integrated communication and localization towards 6G: \cite{xiao2022overview}. }

% Good to read: \cite{khan2020stochastic}

% P5: 
In this paper, we explore the fusion of NTN positioning, more specifically \ac{LEO} positioning, with cellular positioning. The setup to be used comprises one multi-antenna \ac{BS} together with a time and frequency-synchronous single antenna \ac{LEO} satellite transmitting to a single-antenna mobile \ac{UE} which has an unknown clock offset and frequency offset with respect to the \ac{BS} and \ac{LEO} satellite. The objective is to find the position of the \ac{UE} as well as the magnitude of velocity (speed), and clock and frequency offsets.
The contributions of this paper are listed as follows:
\begin{itemize}
    \item \textbf{Discrete-time channel modeling of the integrated \ac{LEO}-cellular network:} We derive a sophisticated generative model based on the integration of BS-UE and satellite-UE communication that accounts for the slow-time and fast-time Doppler, intersubcarrier Doppler effect, \ac{ICI} as well as time-varying \ac{AoD} and path gains resulting from the UE's motion. % based on the coordination of \ac{BS}-\ac{UE} and satellite-\ac{UE},
    The satellite and BS are assumed to be synchronized; but the UE is not, leading to an unknown initial clock offset as well as an unknown frequency offset.
    \item \textbf{Simplified channel modeling based on the derived generative model:} We developed four simplified models based on the derived generative model in increasing order of complexity. The first one is a conventional communication \ac{OFDM} model with a constant phase rotation over the observation window. The second model considers the slow-time Doppler effect on the carrier frequency, where the Doppler is modeled as a phase rotation across OFDM symbols. The third model considers not only the slow-time Doppler effect over the carrier frequency but also over the subcarriers, which leads to intersubcarrier Doppler effect \cite{zhang2020joint}. Finally, our most complex model considers both slow-time and fast-time Doppler effects, which causes \ac{ICI} \cite{prasad2004ofdm,zhang2020joint}.
    \item \textbf{Low complexity positioning and synchronization:} We design low-complexity algorithms for estimating the channel parameters, namely, the propagation delays, Doppler shifts, and \ac{AoD}. These algorithms enable the estimation of the user’s position, velocity magnitude, initial clock offset, and frequency offset based on simplified models.
    \item \textbf{Model Evaluation and Selection:} We assess the simplified models using the \ac{CRB} and a bias metric derived from the \ac{MCRB}, which quantifies the performance degradation caused by using a mismatched estimation model instead of the true generative model. This analysis provides insight into how well the simplified models approximate the generative model, achieving acceptable localization performance while keeping estimation complexity low. These results guide the selection of the most suitable simplified model based on application requirements and deployment scenarios.
    % \item The performance of each simplified  models is analyzed in different scenarios.
\end{itemize}
\textit{Notation:} Vectors and matrices are shown by bold-face lower-case and bold-face upper-case letters respectively. The notations $(.)^\trp$ and $(.)^\her$ represent transpose and hermitian transpose. All one vector with size $n$ denoted by $\boldsymbol{1}_n$. The L2 norm of a vector is shown by $\norm{.}$. The Hadamard product and convolution are represented by $\odot$ and $\ast$ respectively. The set of real numbers and complex numbers are shown by $\mathbb{R}$ and $\mathbb{C}$ respectively. The delta Dirac function is represented by $\delta(.)$. The function $\text{rect}(x)$ is defined such that $\text{rect}(x) = 1$ for $0<x<1$ and is $0$ otherwise. The $\thn{(m,n)}$ element of the matrix $\bA$ is referred to by $[\bA]_{(m,n)}$ and the notation $\tilde{(.)}$ represents a passband signal.
%\begin{table}[]
%    \centering
%    \begin{tabular}{|c|c|c|}
%    \hline
%        Symbol & Text & Units \\
%        \hline
%         & & \\
%         \hline
%    \end{tabular}
%    \caption{Parameters used in the paper}
%    \label{tab_param}
%\end{table}
\vspace{-2mm}
%%%%%%%%%%%%%%%%%%%%%% system model %%%%%%%%%%%%%%%%%%%%%
\section{System Model}
\subsection{Scenario}
We consider a system consisting of a \ac{BS} equipped with a \ac{UPA} of $\sqrt{L}\times \sqrt{L}$
antennas located at known location $\ppbs \in \realsetone{3}$, a \ac{LEO} satellite with a directional antenna with known and varying location and velocity $\ppsat(t)\in\realsetone{3}$ and $\vv_\SAT(t) \in \realsetone{3}$, respectively, and a single-antenna \ac{UE}. The setup is illustrated in Fig. \ref{fig_NTN_setup}. The UE travels at a constant velocity $\vv\in\realsetone{3}$ with known direction $\vv/ \norm{\vv}$ \footnote{The direction can be estimated using inertial sensors (e.g., \acp{IMU}) through dead reckoning or short-term motion tracking, even without reliable GNSS.}, but unknown speed $\norm{\vv}$ and unknown initial location $\pp_0\in\realsetone{3}$. As a result, the trajectory of the \ac{UE} is given by $\pp(t) = \pp_0 + \vv t$.
% Communication occurs using pilots signals over \ac{OFDM}, where the transmission spans $M$ \ac{OFDM} symbols, each comprising $N$ subcarriers.
The clock oscillators at the \ac{BS} and the satellite are assumed to be synchronized to a common global reference. The clock oscillator at the \ac{UE} is assumed to have an unknown initial clock offset, denoted as $\Delta_{t,0}$, relative to the clock oscillators at the \ac{BS} and the satellite. Additionally, the UE's clock oscillator introduces a \ac{CFO} $\eta$ which is treated as an unknown. %, such that if it aims to generate a signal with frequency $f$, the actual generated frequency will be $f(1-\eta)$.

\subsection{Signal and Channel Model}
%In this subsection we formulate the continuous-time band-pass received signal in high level. 
% We assume that the BS and satellite each transmit $M$ OFDM symbols with $N$ subcarriers and subcarrier spacing of $\Delta_f$. 
We assume the positioning task is performed using one transmission consisting $M$ \ac{OFDM} symbols, each comprising $N$ subcarriers (with a subcarrier spacing of $\Delta_f$). We assign the even and the odd subcarriers to the signal transmitted by the \ac{BS} and the satellite, respectively%\cite{dwivedi2021positioning}
. The baseband transmit signal at the base station ($\zz_\BS(t) \in \complexsetone{L}$) is
\begin{align} \label{eq_TX_BS_BB}
 \zz_\BS(t) &= \bw(t) x_\BS(t) = \bw(t) \sum_{m = 0}^{M-1} x_{\BS,m}(t),
 \end{align}
 where
%\begingroup
%\small
\begin{align}
&  x_{\BS,m}(t)  =
 \frac{1}{\sqrt{N}}  \sum_{n=0, \text{even}~n}^{N-1} x_{n,m} e^{j2 \pi n \Delta_f (t-m\Ts)} q(t - m\Ts).\notag 
\end{align}
%\endgroup
Here, $q(t)=\text{rect}(t/\Ts)$,%denotes the transmit pulse shape
\footnote{In case of using a more practical \ac{RRC} pulse instead of a rectangular pulse, we need to apply matched filters at the receiver.}  
$\ww(t) \in \complexsetone{L}$ is the time-varying precoding matrix, $ x_{n,m} $ is the pilot at the $\thn{n}$ subcarrier of the $\thn{m}$ symbol. The total symbol duration is $\Ts = T_0 +\Tcp$, where $T_0 = 1/\Delta_f$ is the elementary symbol duration, and $\Tcp$ is the cyclic prefix duration. Then the transmit signal upconverted to the carrier frequency $f_c$ is $ \tilde\zz_\BS(t) = \Re\{\zz_\BS(t) e^{j2\pi f_c t}\}$.
%\begin{align} \label{eq_TX_BS_BP}
%    \tilde\zz_\BS(t) = \Re\{\zz_\BS(t) e^{j2\pi f_c t}\}.
%\end{align}
% Assuming the \ac{UE} is located in the far-field region of the \ac{BS}, 

The time-varying channel between the BS and the UE $(\tilde\hh_\BS(t, \tau) \in \complexsetone{L})$ can be expressed as follows, for the $\thn{l}$ BS antenna:
\begin{align} \label{eq_hBS_HL}
    \tilde h_{\BS, l}(t, \tau) = [\tilde\hh_\BS(t, \tau)]_l = \alpha_\BS(t) \delta(\tau - \tau_{\BS, l}^p(t)),
\end{align}
where $\alpha_\BS(t)$ denotes the time-varying real-valued channel gain between the BS and the UE and $\tau_{\BS,l}^p(t)$ expresses\footnote{The superscript $"p"$ indicates that the delay is purely due to propagation, distinguishing it from the total effective delay values that will be introduced later.}  the time-varying propagation delay between the $\thn{l}$ antenna at the \ac{BS} and the UE. Therefore, the passband noise-free received signal at the UE can be written as
\begin{align}\label{eq_YBS_cont}
   % \tilde y_\BS(t) =\tilde \bh_\BS^\trp(t, \tau) \ast \tilde \zz_\BS(t).
     %\tilde y_\BS(t) = \sum_{l = 1}^L\tilde h_{\BS,l}(t, \tau) \ast \tilde z_{\BS, l}(t) .
     & \tilde y_\BS(t) = \alpha_\BS(t)
     \sum_{l=1}^L \tilde z_{\BS,l}(t-\tau_{\BS, l}^p(t)).
\end{align}
% The time-varying passband channel between the $l^\text{th}$ antenna at the BS and the UE is denoted by $\tilde h_{\BS,l}(t, \tau) = \alpha_\BS(t)\delta(\tau-\tau_{\BS,l}(t))$, where $\tau_{\BS,l}(t)$ incorporates the effect of delay from the $l^{th}$ BS antenna and $l=0,\cdots,L-1$. Therefore the noise-free received signal to the UE from the BS will be

Similar to the BS-UE transmission, the baseband transmit signal at the satellite is $x_\SAT(t) = \sum_{m = 0}^{M-1} x_{\SAT,m}(t)$, 
%\begin{align} \label{eq_TX_SAT_BB}
%& x_\SAT(t) = \sum_{m = 0}^{M-1} x_{\SAT,m}(t),
%\end{align}
where
\begin{align}
  &  x_{\SAT,m}(t) = 
   \frac{1}{\sqrt{N}}\sum_{n=0, \text{odd}~n}^{N-1} x_{n,m} e^{j2 \pi n \Delta_f (t-m\Ts)} q(t - m\Ts).\notag 
\end{align}
Here, $x_{n,m}$ is the pilot signal transmitted by the satellite. Then the upconverted transmit signal is $\tilde s_\SAT(t) = \Re\{x_\SAT(t) e^{j2\pi f_c t}\}$.
%\begin{align} \label{eq_TX_SAT_BP}
%    \tilde s_\SAT(t) = \Re\{x_\SAT(t) e^{j2\pi f_c t}\}.
%\end{align}
We denote the time-varying channel between the satellite and the UE by 
%\begin{align}
    $\tilde h_{\SAT}(t, \tau) = \alpha_\SAT(t)\delta(\tau-\tau_\SAT^p(t))$, 
%\end{align}
where $\alpha_\SAT(t)$ and $\tau_\SAT^p(t)$ denote the time-varying channel gain and propagation delay between the satellite and the UE. Therefore, the noise-free received signal at the UE from the satellite will be
\begin{align} \label{eq_YSAT_cont}
  & \tilde y_\SAT(t) = \alpha_\SAT(t)
     \tilde s_{\SAT}(t-\tau_{\SAT}^p(t)).
  %  \tilde y_\SAT(t) = \tilde h_{\SAT}(t, \tau) \ast \tilde x_\SAT(t).
\end{align}
Finally, the total passband received signal will be
\begin{align} \label{eq_yt}
    \tilde y(t) = \tilde y_\BS(t) + \tilde y_\SAT(t) + \tilde n(t),
\end{align}
where $\tilde n(t)$ is the passband \ac{AWGN} at the UE, with power spectral density $N_0$.
% - abstract - start from TX signal from BS and satellite, orthogonal subcarriers, ... , no precompensation. For the observation Model: in an abstract way write $y(t) = h_\BS(t) \ast s_\BS(t) + h_\SAT(t) \ast s_\SAT(t) + n(t)$, so in avery abstract way h(t, tau), and h of BS and SAT. the passband representation as $h_{\BS,i}(t) = \alpha(t) \delta(t-\tau_i(t))$ where $\tau_i(t)$ incorporates the effect of delay from the $i^{th}$ BS antenna. Similar for $h_\SAT(t)$.

\subsection{Geometric Relations}

The time-varying propagation delay between the $\thn{l}$ antenna of the \ac{BS} and the \ac{UE} is given by
\begin{align}
    \tau_{\BS,l}^p(t) = \tau_\BS^p(t) + \tau_l^p(t),
\end{align}
where $\tau_\BS^p(t) = \norm{\pp(t) - \ppbs}/c$ is the delay between the \ac{BS} phase center and the \ac{UE}, and $c$ is the speed of light. Assuming the \ac{UE} is located in the far-field region of the \ac{BS}, the relative delay (or advance) of the $\thn{l}$ antenna element with respect to the phase center is  $\tau_l^p(t) = -( \uu(\thetab(t))^\trp \qq_l)/c$, where $\qq_l \in \realsetone{3}$ is the known position of the $\thn{l}$ antenna element with respect to the \ac{BS} phase center. %, expressed in the local coordinate system of the \ac{BS}.
The unit direction vector $\uu(\thetab(t)) \in \realsetone{3}$ captures the orientation of the \ac{UE} with respect to the \ac{BS}, and is defined as
 $\uu(\thetab(t)) =  \left[ \cos(\thetael(t)) \cos(\thetaaz(t)), \, \cos(\thetael(t)) \sin(\thetaaz(t)), \, \sin(\thetael(t)) \right]^\trp$,
where $\thetab(t) = \big[\thetael(t), \thetaaz(t) \big]^\trp \in \realsetone{2}$ is the time-varying 2D \ac{AoD} in the global coordinate system. These angles are computed based on the relative position vector $\rr_\BS(t) = \pp(t) - \ppbs$ from the \ac{BS} to the \ac{UE}, as
\begin{align} \label{eq_geo_rel_AoD}
    \thetael(t) &= \arcsin\left( \frac{[\rr_\BS(t)]_3}{\norm{\rr_\BS(t)}} \right), \\
    \thetaaz(t) &= \atant\left( [\rr_\BS(t)]_2, [\rr_\BS(t)]_1 \right).
\end{align}

Finally, the time-varying propagation delay between the satellite and the \ac{UE} is given by
%\begin{align}
    $\tau_\SAT^p(t) = {\norm{\pp(t) - \ppsat(t)}}/{c}$.
%\end{align}

\subsection{Problem Statement}
The UE should determine its initial position $\pp_0$ and its speed $\norm{\vv}$, from the observation $\tilde y(t)$ in \eqref{eq_yt}, in the presence of the unknown and time-varying channel parameters, and without being a priori synchronized to the network (the BS and the satellite). The details of the UE-network asynchrony will be elaborated in the next section.
 For ease of reference, the parameters introduced in the following sections are summarized in Table~\ref{tab:parameters}.

\section{Generative and Simplified Models}\label{sec_gen_simp_models}
In this section, we begin by introducing the notion of time-varying clock bias; then, we derive the generative model used to generate observations for the localization task. To assist algorithm development, we propose four distinct simplified models, some of which are widely used in the literature.
% the proposed methods in Section \ref{sec_methods}. These methods are derived based on 4 distinct mismatched models, some of which are widely used in the literature. 

\subsection{Clock and Frequency Offset}
Due to the imperfect clock oscillator at the UE, the notion of time at the UE differs from that at the BS and satellite. The time reference of the receiver is denoted by $t'$ and is supposed to be related to the time reference at the transmitter through 
\begin{align} \label{eq_t'}
    t' = t/(1-\eta) + \Delta_{t,0},
\end{align}
where $t$ represents the network reference time and $\Delta_{t,0}$ is the initial clock offset. Hence, for $\abs{\eta}\ll 1$ it holds that the time-varying offset $\Delta_t(t)= t' - t$ can be expressed as 
%\begin{align}
    $\Delta_t(t) \approx \eta t + \Delta_{t,0}$~\cite{riley2008handbook}.
\subsection{Generative Model}
\subsubsection{Continuous-time Model}
% Let's explore $\tilde\bh_\BS(t, \tau)$ \eqref{eq_hBS_HL} in more detail. 
%The channel between the $\thn{l}$ BS antenna and the \ac{UE} is
%\begin{align}
%    \tilde h_{\BS,l}(t, \tau) = \alpha_\BS(t) [\aab(\thetab(t))]_l \delta(\tau-\tau_\BS^p(t)),
%\end{align}

The received passband signal in the global time reference $t$ in \eqref{eq_YBS_cont} and \eqref{eq_YSAT_cont} has the below contributions from the BS-UE path and satellite-UE path (see Appendix \ref{eq_genmodel_deriv}): 
\begin{align}
& \tilde{y}_\BS(t) =\\
&  \Re\{ \alpha_\BS(t) \aab^\trp(\thetab(t)) \ww(t - \tau_\BS^p(t)) x_\BS\left(t - \tau_\BS^p(t)\right) %\nonumber \\ & 
e^{j 2 \pi f_c \left(t  - \tau_\BS^p(t)\right)}\} \notag  
%\\
%& = \Re\big\{ \alpha_\BS(t) \aab^\trp(\thetab(t)) \ww(t - \tau_\BS^p(t)) \nonumber \\ & \sum_{m = 0}^{M-1}  \frac{1}{\sqrt{N}} \sum_{n=0}^{N-1} x_{n,m}  e^{j 2 \pi f_c \left(t  - \tau_\BS^p(t) \right)} \nonumber \\& \times e^{j 2 \pi n\Delta_f \left(t  - \tau_\BS^p(t)- m\Ts \right)} q({t - \tau_\BS^p(t) - m\Ts})\big\} ,
\end{align}
where the array steering vector $\aab(\thetab(t))$ is given by 
%\begin{align}
    $[\aab(\thetab(t))]_l = \exp( j ({2\pi}/{\lambda}) \uu^\trp(\thetab(t)) \qq_l )$ \cite{Kamran_JSTSP_SISO_RIS},
%\end{align} 
and 
\begin{align}
    & \tilde{y}_\SAT(t) = \Re\big\{ \alpha_\SAT(t) x_\SAT\left(t - \tau_\SAT^p(t)\right) e^{j 2 \pi f_c \left(t  - \tau_\SAT^p(t)\right)}\big\}. %\nonumber \\ =& \Re\bigg\{ \alpha_\SAT(t) \sum_{m = 0}^{M-1} \frac{1}{\sqrt{N}} \sum_{n=0}^{N-1} x_{n,m}  e^{j 2 \pi n\Delta_f \left(t  - \tau_\SAT^p(t)-m\Ts\right)} \nonumber \\ & \times e^{j 2 \pi f_c \left(t  - \tau_\SAT^p(t)\right)}\text{rect}\left(\frac{t - \tau_\SAT^p(t) - m\Ts}{\Ts}\right)\bigg\}.
\end{align}
%Since the UE's velocity is assumed to be constant (therefore \eqref{eq_smallDisplacement} holds), 
% The time-varying propagation delay between the BS and the UE and satellite and the UE are denoted by $\tau_\BS^p(t)$ and $\tau_\SAT^p(t)$ respectively and are defined and approximated as in below:
It is possible to approximate $\tau_\Ent^p(t)$, (with $\Ent \in \{\BS, \SAT\}$) as (see Appendix \ref{app_lin_tau})
\begin{align}
    %\begin{split} 
    \label{eq_tau_BS_SAT_quad}
    \tau_\Ent^p(t) & = \frac{\norm{\pp(t) - \pp_\Ent(t)}}{c}
     \approx \frac{\norm{\pp_0 - \pp_\Ent}}{c} + \frac{v_{\Ent,\UE}}{c}t+\frac{a_{\Ent,\UE}}{2c} t^2\nonumber \\
     & = \tau_\Ent^p + \psi_\Ent t + \frac{1}{2} \mu_\Ent t^2,
  %  \tau_\BS^p(t) & = \frac{\norm{\pp(t) - \ppbs}}{c}
   %  \approx \frac{\norm{\pp_0 - \ppbs}}{c} + \frac{v_{\bu, 0}}{c}t+\frac{a_\bu}{2c} t^2\nonumber \\
   %  & = \tau_\BS^p + \psi_\BS t + \frac{1}{2} \mu_\BS t^2 \\
    %\tau_\SAT^p(t) & = \frac{\norm{\pp(t) - \ppsat(t)}}{c} \approx \frac{\norm{\pp_0 - \ppsat}}{c} + \frac{v_{\su, 0}}{c}t+\frac{a_\su}{2c} t^2\nonumber \\
    % & = \tau_\SAT^p + \psi_\SAT t + \frac{1}{2} \mu_\SAT t^2, \label{eq_tau_SAT_quad}
    %\end{split}
\end{align}
and $\tau_\Ent^p = \norm{\pp_0 - \pp_\Ent}/c$, $\psi_\Ent  = {v_{\Ent,\UE}}/{c}$,  $ v_{\Ent,\UE}  = (\pp_0 - \pp_\Ent)^\trp \vv_{\Ent,\UE}/\norm{\pp_0 - \pp_\Ent}$, $ \mu_\Ent  = a_{\Ent,\UE}/c $, and $a_{\Ent,\UE}  = (\norm{\vv_{\Ent,\UE}}^2-v_{\Ent,\UE}^2)/\norm{\pp_0 - \pp_\Ent}$, 
%\begin{subequations}
%    \begin{align}
%& \tau_\Ent = \norm{\pp_0 - \pp_\Ent}/c, \\
% &   \psi_\Ent  = {v_{\Ent,\UE}}/{c}, \label{eq_psi_BS_SAT}\\
%  &  v_{\Ent,\UE}  = (\pp_0 - \pp_\Ent)^\trp \vv_{\Ent,\UE}/\norm{\pp_0 - \pp_\Ent}, \label{eq_v_bu0}\\
%   & \mu_\Ent  = a_{\Ent,\UE}/c ~, \\ 
 %   &a_{\Ent,\UE}  = \norm{\vv_{\Ent,\UE}}^2/\norm{\pp_0 - \pp_\Ent}. %\\ 
%  \tau_\BS^p & = \norm{\pp_0 - \ppbs}/c, \\
 %  \psi_\BS & = {v_{\bu, 0}}/{c}, \label{eq_psi_BS}\\
%    v_{\bu, 0} & = (\pp_0 - \ppbs)^\trp \vv/\norm{\pp_0 - \ppbs}, \label{eq_v_bu0}\\
%    \mu_\BS  & = a_\bu/c ~, a_\bu = \norm{\vv}^2/\norm{\pp_0 - \ppbs}. \\ 
%    \tau_\SAT^p & = \norm{\pp_0 - \ppsato}/c, \\
%   \psi_\SAT & = {v_{\su, 0}}/{c}, \label{eq_psi_SAT}\\
%    v_{\su, 0} & = (\pp_0 - \ppsato)^\trp \vv_\su/\norm{\pp_0 - \ppsato}, \label{eq_v_su0}\\
 %   \vv_\su & = \vv_\LEO - (\vv_\E + \vv), \label{eq_vv_su}\\
 %   \mu_\SAT  & = a_\su/c ~, a_\su = \norm{\vv_\su}^2/\norm{\pp_0 - \ppsato}.
%\end{align}
%\end{subequations}
in which $\vv_{\BS,\UE} = \vv$ and $\vv_{\SAT,\UE} =
(\vv + \vv_\E)-\vv_\LEO $, where $\vv_\E$ denotes the Earth rotational velocity.

% The details on the approximation in \eqref{eq_tau_BS_SAT_quad} %and \eqref{eq_tau_SAT_quad} 
% are elaborated in Appendix \ref{app_lin_tau}.
%\subsection{True Model} 
% So far, the received signal is written in transmitter's time domain.
We now rewrite the received signal \textit{in the receiver's time domain} $t'$ by substituting $t = (t' - \Delta_{t,0})(1-\eta)$ according to \eqref{eq_t'} and using \eqref{eq_tau_BS_SAT_quad}
\begin{comment}
\begin{align} \label{eq_tilde_y_t'}
    &\tilde y'(t') = \tilde y'_\BS(t') + \tilde y'_\SAT(t') + \tilde n'(t')  \nonumber \\ & = \Re\big\{\alpha'_\BS(t') \aab'^\trp(\thetab'(t')) \ww'(t' - {\tau'_b}^p(t')) x'_\BS\left(t' - {\tau'_\BS}^p(t')\right) \nonumber \\ &  \quad  e^{j2\pi f_c \left(t' - {\tau'_b}^p(t')\right)} \nonumber \\ & + \alpha'_\SAT(t') x'_\SAT\left(t' - {\tau'_\SAT}^p(t')\right) e^{j2\pi f_c \left( t' - {\tau'_\SAT}^p(t')\right)}  + \tilde n'(t') \big\},
\end{align}
\end{comment}
\begin{align} \label{eq_tilde_y_t'}
    \tilde y'(t') & = \tilde y'_\BS(t') + \tilde y'_\SAT(t') + \tilde n'(t')  \nonumber \\ & = \Re\big\{\alpha'_\BS(t') \aab'^\trp(\thetab'(t')) \ww'(t_b) x'_\BS\left(t_b\right)  e^{j2\pi f_c t_b} \nonumber \\ & + \alpha'_\SAT(t') x'_\SAT\left(t_s\right) e^{j2\pi f_c  t_s}  + \tilde n'(t') \big\},
\end{align}
\begin{comment}
\begin{align} \label{eq_tilde_y_t'}
    &\tilde y'(t') = \tilde y'_\BS(t') + \tilde y'_\SAT(t') + \tilde n'(t')  \nonumber \\ & = \Re\big\{\alpha'_\BS(t') \aab'^\trp(\thetab'(t')) \ww'(t_b)  \sum_{m = 0}^{M-1}  \frac{1}{\sqrt{N}} \sum_{n=0, ~\text{even n}}^{N-1} x_{n,m} \nonumber \\ &  \quad  e^{j2 \pi n \Delta_f \left(t_b - m\Ts \right)}  e^{j2\pi f_c t_b} q({t_b - m\Ts })  \nonumber \\ & + \alpha'_\SAT(t') \sum_{m = 0}^{M-1} \frac{1}{\sqrt{N}} \sum_{n=0, ~\text{odd n}}^{N-1} x_{n,m}   e^{j2 \pi n \Delta_f \left(t_s  - m\Ts \right)} \nonumber \\ & \quad e^{j2\pi f_c t_s} q({t_s - m\Ts })  + \tilde n'(t') \big\},
\end{align}
\end{comment}
where $\alpha'_\BS(t') = \alpha_\BS((t'-\Delta_{t,0})(1-\eta))$, $\thetab'(t') = \thetab((t' - \Delta_{t,0})(1-\eta))$. To ease the notation, we have introduced $t_b = (1-\gamma_b)t' - \tau_b - \epsilon_b t'^2$, $t_s = (1-\gamma_s)t' - \tau_s - \epsilon_s t'^2$ and define $\bw'(t_b) = \bw((1-\gamma_b)t' - \tau_b - \epsilon_b t'^2)$, % $\bw'(t_s) = \bw((1-\gamma_s)t' - \tau_s - \epsilon_s t'^2)$,
and
\begin{subequations}
\begin{align}
 \gamma_\Ent & = {\psi_\Ent}({1-\eta}) +\eta -{\mu_\Ent}(1-\eta)^2\Delta_{t,0}, \label{eq_gamma_b}\\
    \epsilon_\Ent & = 1/2\mu_\Ent(1-\eta)^2, \\ \tau_\Ent & = \tau_\Ent^p  + \Delta_{t,0}(1-\eta)(1-\psi_\Ent) + 1/2 \mu_\Ent \Delta_{t,0}^2(1-\eta)^2, \label{eq_tau_i}
    %\gamma_b & = {\psi_\BS}({1-\eta}) +\eta -{\mu_\BS}(1-\eta)^2\Delta_{t,0}, \label{eq_gamma_b}\\
    %\epsilon_b & = \mu_\BS/2(1-\eta)^2, \\ \tau_b & = \tau_\BS^p  + \Delta_{t,0}(1-\eta)(1-\psi_\BS) + 1/2 \mu_\BS \Delta_{t,0}^2(1-\eta)^2, \\
    %\gamma_s & = \psi_\SAT({1-\eta}) + \eta -{\mu_\SAT}(1-\eta)^2\Delta_{t,0}, \label{eq_gamma_s}\\
    %\epsilon_s & = \mu_\SAT/2(1-\eta)^2, \\
    % \tau_s & = \tau_\SAT^p  + \Delta_{t,0}(1-\eta)(1-\psi_\SAT) +  1/2 \mu_\SAT \Delta_{t,0}^2(1-\eta)^2.
\end{align}
\end{subequations}
where $\Ent \in \{b, s\}$. Here, $\gamma_\Ent$, $\epsilon_\Ent$, and $\tau_\Ent$ can be interpreted as the effective Doppler shift, effective Doppler rate, and effective delay, respectively, in a system affected by clock offset, carrier frequency offset, and time-varying Doppler. This completes the continuous-time generative model, accounting for both first- and second-order Doppler effects as well as the asynchrony between the \ac{UE} and the network.

\begin{table}%[t]
\centering
\caption{List of Parameters}
\label{tab:parameters}
\begin{tabular}{ll}
\hline
\textbf{Symbol} & \textbf{Description} \\
\hline
$\pp_0$ & Initial position of the UE \\
$\vv$ & Velocity vector of the UE \\
$\norm{\vv}$ & Speed of the UE \\
$\pp(t)$ & Position of the UE at time $t$ \\
$\ppbs$ & Position of the BS \\
$\ppsat(t)$ & Position of the satellite at time $t$ \\
$\vv_{\SAT}$ & Velocity of the satellite \\
$\vv_E$ & Earth's rotational velocity \\
$\Ts$ & Total symbol duration \\
$\Tcp$ & Cyclic prefix duration \\
$T_0$ & Elementary symbol duration \\
$\Delta_f$ & Subcarrier spacing \\
$\qq_l$ & Position of the $\thn{l}$  BS antenna element \\
$\thetab(t)$ & 2D AoD (elevation, azimuth) at time $t$, $[\thetael(t), \thetaaz(t)]^\trp$ \\
$\rr_{\BS}(t)$ & Vector from the BS to the UE \\
$\tau_{\BS,l}^p(t)$ & Propagation delay from $\thn{l}$  BS antenna to UE \\
$\tau_{\BS}^p(t)$ & Propagation delay from BS center to UE \\
$\tau_l^p(t)$ & Relative delay from $\thn{l}$  antenna to BS center \\
$\tau_{\SAT}^p(t)$ & Propagation delay from satellite to UE \\
$\Delta_{t,0}$ & Initial clock offset of the UE \\
$\eta$ & CFO of the UE \\
$t$ & Global (network) reference time \\
$t'$ & Local time at the UE \\
$\Delta_t(t)$ & Time-varying clock offset \\
$\tau_i^p(t)$ & Propagation delay from node $i \in \{\BS, \SAT\}$ to UE \\
$\tau_i^p$ & Initial propagation delay from node $i \in \{\BS, \SAT\}$ to UE \\
$\psi_i$ & First-order Doppler term for node $i$ \\
$\mu_i$ & Second-order Doppler term for node $i$ \\
$\gamma_i$ & Effective Doppler shift for node $i$ \\
$\epsilon_i$ & Effective Doppler rate for node $i$ \\
$\tau_i$ & Effective initial delay for node $i$ \\
$\bar{\psi}_{\bs}$ & Proxy Doppler term for satellite-BS path \\
$\vv_{\bs}$ & Relative velocity between BS and satellite \\
$\tau_{\bs}$ & Estimated delay between BS and satellite \\
$\tau_s^{\mathrm{res}}$ & Initial residual delay after satellite timing advance \\
\hline
\end{tabular}
\end{table}

\subsubsection{Doppler and Time Precompensation}
Based on \eqref{eq_tilde_y_t'}, the received frequency through the BS-UE path is $f_c (1-\gamma_b)$ and through the satellite-UE path is $f_c (1-\gamma_s)$. This results in a significant Doppler spread of the received signal due to high velocity of the LEO satellite, which causes limitations in low-pass filtering prior to sampling at the receiver. Although $\gamma_b$ is unknown and much smaller than $\gamma_s$, a significant portion of $\gamma_s$ can be determined from the known satellite velocity and position, as well as the known BS position, under the assumption that the UE is in close proximity to the BS. Consequently, it is feasible to apply a frequency (Doppler) precompensation at the satellite prior to transmission, thereby rendering the Doppler shifts in the received signals through different paths more comparable and reducing the overall Doppler spread. To achieve this, a proxy for $\gamma_s$, denoted by $\bar \psi_{\bs}$, can be estimated as follows (where the subscript `$\bs$' represents satellite-BS path):
\begin{align}
    \bar \psi_{\bs} & = v_{\bs}/c, \\
    v_{\bs} & =  (\ppbs - \ppsato)^\trp \vv_\bs/\norm{\ppbs - \ppsato}, \\
    \vv_\bs & = \vv_\E -  \vv_\LEO .
\end{align}
Then, we can utilize $ \bar \psi_{\bs}$ to adjust the transmit frequency at the satellite to $\bar {f_c} = f_c/(1- \bar \psi_{\bs})$. With this Doppler precompensation, the received frequency in the satellite-UE path is changed from $f_c(1-\gamma_s)$ to $f_c(1-\gamma_s)/(1-\bar\psi_\bs)$.

Similarly, in the time domain, the propagation delay experienced by the signal in the satellite-UE path is considerably larger than that in the BS-UE path. Nevertheless, a significant portion of the satellite-UE propagation delay is already known from the satellite and BS positions, as well as the UE’s proximity to the BS. Hence, it is possible to apply timing advance at the satellite to reduce the delay spread in the received signal. Analogous to Doppler precompensation, a rough estimate of $\tau_\SAT^p$ can be obtained as 
\begin{align}
    \tau_\bs = \norm{\ppbs - \ppsato}/c,
\end{align}
and we can then apply a timing advance of $\tau_\bs$ to the satellite transmit signal $s(t)$, expressed as 
\begin{align}
     \tilde {\bar  {s}}_\SAT(t) = \Re\{x_\SAT(t+\tau_\bs) e^{j 2\pi f_c (t+\tau_\bs)}\}.
\end{align}
In this case, $\tau_s$ is replaced by $\tau_s^\res$ \footnote{The superscript 'res' stands for residual.} in \eqref{eq_tilde_y_t'} defined as follows:
\begin{align}
    \tau_s^\res &= \tau_\SAT - \tau_\bs.
\end{align}
% In this case, the total frequency spread in the received signal changes from $f_c\bigg((\psi_\SAT - \psi_\BS)(1-\eta) - (\mu_\SAT - \mu_\BS)(1-\eta)^2\Delta_{t,0}\bigg)$ to $f_c\bigg((\psi_\SAT - \psi_\BS)(1-\eta) - (\mu_\SAT - \mu_\BS)(1-\eta)^2\Delta_{t,0}\bigg)$ 
\subsubsection{Down-Conversion}
To down-convert to the baseband, the clock oscillator at the UE generates $e^{-j2 \pi f_c t'}$. Consequently, after removing the constant phases, the baseband signal becomes:
\begin{align} \label{eq_y_t'}
      & y'(t') =\tilde y'(t') e^{-j2 \pi f_c t'} =\alpha'_\BS(t') z(t')  \sum_{m = 0}^{M-1}  \frac{1}{\sqrt{N}} \sum_{n=0, ~\text{even n}}^{N-1} \nonumber \\ & x_{n,m} e^{j2 \pi n \Delta_f \left(t_b  - m\Ts \right)}  e^{-j2\pi f_c \left( \gamma_b t' + \epsilon_b t'^2\right)} q({t_b - m\Ts })  \nonumber \\ & + \alpha'_\SAT(t') \sum_{m = 0}^{M-1} \frac{1}{\sqrt{N}} \sum_{n=0, ~\text{odd n}}^{N-1} x_{n,m}e^{j2 \pi n \Delta_f \left(t_s^\res - m\Ts \right)} \nonumber \\ & \times  e^{-j2\pi \bar f_c \left( \gamma_s t' + \epsilon_s t'^2\right)}  e^{j2 \pi (\bar f_c - f_c)t'}  q({t_s^\res - m\Ts })  +  n'(t'),
\end{align}
where $z(t') = \aab^\trp(\thetab(t'))\ww'(t_b)$ and $t_s^\res = (1-\gamma_s)t' - \tau_s^\res - \epsilon_s t'^2$.
\subsubsection{Discrete-time Model}
To obtain the discrete-time received signal, we first apply analog-domain low-pass filtering to suppress out-of-band components, and then sample~\eqref{eq_y_t'} at time instants \( t' = m\Ts + \Tcp + kT_0/N + \tau_0 \), where \( k = 0, 1, \dots, N-1 \), and
\begin{align}\label{eq_tau0}
    \tau_0 = \min\left({\tau_\BS}/(1-\eta), {\tau_\SAT}/(1-\eta)\right),
\end{align}
represents the smallest delay in the receiver's time domain that ensures the first OFDM symbol is fully captured. In~\eqref{eq_tau0}, \( \tau_\BS \) and \( \tau_\SAT \) are defined in~\eqref{eq_tau_i}. Let us assume that $\tau_0$ is detectable (through time acquisition in \cite{etsi2020138}) and hence we can start the receiver's clock at $\tau_0$. 
We can take $N$ samples from each OFDM symbol, resulting in $\bY  = [\by_0, \cdots, \by_{M-1}] \in \mathbb{C}^{N \times M}$ where $\by_m$ is defined as
\begin{align} \label{eq_TM}
    &\by_m = \by_{\BS,m}+ \by_{\SAT,m} + \bn_m,
\end{align}
    with
\begin{align}
     \by_{\BS,m} & =   
    \sqrt{P_{\BS}} [\bA_\BS]_{(:,m)}\odot [\bZ]_{(:,m)} \odot \bigg(\bD_{\BS}^q (\gamma_b, \epsilon_b) \nonumber \\ &  \times  \bF^q_{m, \BS}(\gamma_b, \epsilon_b)  \Big(\bi^q_{m,\BS}(\gamma_b, \epsilon_b) \odot \Big(\big(\bq_{\BS}(\gamma_b) \odot \bb_{\BS}(\tau_b)\big) \nonumber \\ & \times [\bc^q_{\BS}(\gamma_b, \epsilon_b)]_m\Big)\Big)\bigg)   \odot \bH^q_\BS(\epsilon_b),
\end{align}
\begin{align}
     &\by_{\SAT,m} = \sqrt{P_{\SAT}} [\bA_\SAT]_{(:,m)} \odot  \bigg(\bD^q_{\SAT}(\gamma_s, \epsilon_s) \bF^q_{\SAT}(\gamma_s, \epsilon_s) \times  \\  &\Big(\bi^q_{\SAT}(\gamma_s, \epsilon_s) \odot \Big(\big(\bq_{\SAT}(\gamma_s) \odot \bb_{\SAT}(\tau_s^\res)\big) [\bc^q_{\SAT}(\gamma_s, \epsilon_s)]_m\Big) \Big)\bigg)\odot \bH^q_\SAT(\epsilon_s).\nonumber
\end{align}
The superscript `$q$' is used to underscore that the corresponding matrices/vectors contain quadratic terms, which originate from \eqref{eq_tau_BS_SAT_quad}. %and \eqref{eq_tau_SAT_quad}.
The different matrices defined above are used to capture the different effects: \vspace{-2mm}
\begin{itemize}
    \item \textit{Time variations:} 
$[\bA_\BS]_{(:,m)}$, $[\bA_\SAT]_{(:,m)}$ and $[\bZ]_{(:,m)}$ comprise the samples of $\alpha'_\BS(t')$, $\alpha'_\SAT(t')$ and $z(t')$, respectively. % at $t' = m\Ts + \Tcp + k T_0/N, ~ k = 0,\cdots, N-1 $ are located.
\item \textit{Inter-carrier interference:}
The diagonal matrices $\bD_\BS^q(\gamma_b, \epsilon_b) \in \complexset{N}{N}$ and $\bD^q_\SAT(\gamma_s, \epsilon_s)\in \complexset{N}{N}$ denote \ac{ICI} in the BS-UE and satellite-UE paths, respectively. The $\thn{n}$ diagonal element of each matrix % where $n = 0,1,\cdots,N-1$ 
is defined as
%\begingroup
%\small
\begin{align}
    &[\bD_\BS^q(\gamma_b, \epsilon_b)]_{k,k} =\\
    & e^{-j2\pi f_c (\gamma_b {T_0}k/{N} + \epsilon_b T_0^2 k^2/N^2  + 2 \epsilon_b \Tcp T_0 k/N )}, \notag \\
   %   &\bD_\BS^q(\gamma_b, \epsilon_b) =\operatorname{diag}\bigg(\bigg[1, e^{-j2\pi f_c (\gamma {T_0}/{N} + \epsilon_b T_0^2 /N^2  + 2 \epsilon_b \Tcp T_0/N )},\nonumber \\ & \cdots, e^{-j2\pi f_c (\gamma {T_0(N-1)}/{N}  + \epsilon_b  T_0^2 (N-1)^2/N^2+ 2 \epsilon_b \Tcp T_0(N-1)/N )}\bigg]\bigg),
%\end{align}
%\endgroup
%%\begingroup
%\small
% \begin{align}
 & [\bD^q_\SAT(\gamma_s, \epsilon_s)]_{k,k} = 
  e^{-j2\pi f_c(1- ({1-\gamma_s})/({1-\bar \psi_\bs}) ){T_0} k/{N} }  \\  & \times  e^{-j 2\pi f_c (\epsilon_s   T_0^2/(1-\bar\psi_\bs) k^2/N^2 + 2 \epsilon_s \Tcp T_0 /(1-\bar\psi_\bs)k/N )}.\notag 
 %   &\bD^q_\SAT(\gamma_s, \epsilon_s)=  \operatorname{diag}\bigg(\bigg[ \\ & 1, e^{-j2\pi f_c((1- ({1-\gamma})/({1-\tilde \psi_\SAT}) ){T_0}/{N} +\epsilon_s T_0^2 /N^2  + 2 \epsilon_s \Tcp T_0/N )}, \cdots, \nonumber \\& e^{-j2\pi f_c((1- ({1-\gamma})/({1-\tilde \psi_\SAT}) ){T_0}(N-1)/{N} +\epsilon_s T_0^2 (N-1)^2/N^2  + 2 \epsilon_s \Tcp T_0(N-1)/N )}\bigg]\bigg). \nonumber
\end{align}
%\endgroup
\item \textit{Slow-time Doppler: }The slow-time effect in BS-UE path and satellite-UE path are denoted by $\bc^q_\BS(\gamma_b, \epsilon_b)$ and $\bc^q_\SAT(\gamma_s, \epsilon_s)$ respectively and defined as
    \begin{align}
    &[\bc^q_\BS(\gamma_b, \epsilon_b)]_m =  e^{-j2\pi f_c (\gamma m \Ts + \epsilon_b m^2 \Ts^2 + 2\epsilon_b m \Ts \Tcp)}, 
    \end{align}
    \begin{align}
    [\bc^q_\SAT&(\gamma_s, \epsilon_s)]_m = e^{-j2\pi f_c((1-(1-\gamma_s)/(1-\bar\psi_\bs))m\Ts)} \nonumber \\ & \times e^{ -j2\pi f_c( \epsilon_s m^2 \Ts^2/(1-\bar\psi_\bs)+2\epsilon_s m\Ts \Tcp/(1-\bar\psi_\bs))}.
\end{align}

\item \textit{Delay steering vectors:} For even $n$, $\bb_\BS(\tau) \in \complexsetone{N/2}$ with \begin{align}
\bb_\BS(\tau_b) = &  [1, e^{-j2\pi \Delta_f 2\tau_b}, \cdots e^{-j2\pi \Delta_f(N-2)\tau_b}]^\trp.
\end{align} and for odd $n$, \begin{align}
& \bb_\SAT(\tau_s^\res) = \\
&  [ e^{-j2\pi \Delta_f \tau_s^\res}, e^{-j2\pi \Delta_f 3\tau_s^\res}, \cdots e^{-j2\pi \Delta_f(N-1)\tau_s^\res}]^\trp.\notag 
\end{align} 
\item \textit{Modified Fourier matrices:} For even $n$, the matrix $\bF^q_{m, \BS}(\gamma_b, \epsilon_b)\in \complexset{N}{N/2}$ is defined as 
\begin{align}
   &[\bF^q_{m, \BS}(\gamma_b, \epsilon_b)]_{k,n} =  \frac{1}{\sqrt{N}} \times  \\ & e^{j 2\pi (k/Nn(1-\gamma_b)-\epsilon_b k^2/N^2 n T_0- 2\epsilon_b g k/N n T_0 - 2nm\epsilon_bk/N\Ts)}, \nonumber
\end{align} while for odd $n$
\begin{align} 
\label{eq_Fq_SAT}
   & [\bF^q_{m, \SAT}(\gamma_s, \epsilon_s)]_{k,n} = \frac{1}{\sqrt{N}} \times  \\ & e^{j 2\pi (k/Nn(1-\gamma_s)-\epsilon_s k^2/N^2 n T_0- 2\epsilon_s g k/N n T_0 - 2nm\epsilon_s k/N\Ts)}.\nonumber
\end{align}
\item  \textit{Intersubcarrier Doppler effect:} For even $n$, $\bI^q_\BS(\gamma_b, \epsilon_b) \in \complexset{N/2}{M}$, can be expressed as
\begin{align}
    & [\bI^q_\BS(\gamma_b, \epsilon_b)]_{n,m} = \\
    & e^{-j 2\pi ( \gamma_b mn (g+1) + nm^2 \epsilon_b (1+g) \Ts +2mn \epsilon_b g \Ts) }, \notag 
\end{align} while for odd $n$
\begin{align}
 & [\bI^q_{\SAT, m}(\gamma_s, \epsilon_s)]_{n,m} =  \\ & e^{-j 2\pi ( \gamma_s mn (g+1) + nm^2 \epsilon_s (1+g) \Ts +2mn \epsilon_s g \Ts) }.\nonumber
\end{align}
\item \textit{Intersubcarrier phase offset:} For even $n$, $\bq_\BS(\gamma_b)\in \complexsetone{N/2}$, where \begin{align}
    \bq_\BS(\gamma_b) & = [1 , e^{-j 2\pi 
 \gamma_b g 2}, \cdots, e^{-j 2\pi 
 \gamma_b g (N-2)}]^\trp,
 \end{align} while for odd $n$ \begin{align}
\bq_\SAT(\gamma_s) & = [e^{-j 2\pi 
 \gamma_s g }, \cdots, e^{-j 2\pi 
 \gamma_s g (N-1)}]^\trp.
\end{align}
These offsets remain constant across $M$ OFDM symbols.
\item \textit{Second-order crossed slow/fast-time Doppler effect}:
Finally,
\begin{align}
    [\bH^q_\BS(\epsilon_b)]_{k,m} = e^{-j4 \pi f_c \epsilon_b m\Ts kT_0/N},
\end{align}
\begin{align}
    [\bH^q_\SAT(\epsilon_s)]_{k,m} = e^{-j4 \pi f_c/(1-\bar\psi_\bs) \epsilon_s m\Ts kT_0/N}.
\end{align}
\end{itemize}
The received signal in \eqref{eq_TM} represents the full generative model, which is highly complex and challenging to handle. In the following subsection, four simplified models are proposed, arranged in decreasing order of complexity.
\subsection{Simplified Models} \label{subsec_simpmodels}
In this subsection, four simplified models are introduced based on the true model \eqref{eq_TM}, arranged in descending order of complexity.
%\subsubsection{Model D}
\subsubsection{Model with constant channel (gain and AoD) and only first-order Doppler (CCFOD)}
This model is derived by neglecting the second-order terms in \eqref{eq_TM}, and with the assumption that the channel gains and AoDs remain constant throughout the transmission of $M$ OFDM symbols. Specifically, we set $\mu_\BS = 0$ and $\mu_\SAT = 0$, which would change $\gamma_b$, $\gamma_s$, $\tau_\BS$ and $\tau_\SAT^\res$ in \eqref{eq_gamma_b}  and \eqref{eq_tau_i}
 as 
 \begin{subequations}
\begin{align}
    \gamma_b &= \psi_\BS(1-\eta) + \eta, \\
    \gamma_s &= \psi_\SAT(1-\eta) + \eta, \\
    \tau_\BS & = \tau_\BS^p  + \Delta_{t,0}(1-\eta)(1-\psi_\BS), \label{eq_tau_b_1st}\\
    \tau_\SAT & = \tau_\SAT^p  + \Delta_{t,0}(1-\eta)(1-\psi_\SAT). \label{eq_tau_s_1st}
\end{align}
 \end{subequations}
Moreover, we consider $\alpha_\BS = \alpha_\BS(t=0)$, $\alpha_\SAT = \alpha_\SAT(t=0)$, $\thetab = \thetab(t=0)$. Under these assumptions, the received signal simplifies to 
 \begin{align} 
    &\bY =\label{eq_model_d} \\ & \sqrt{P_\BS}\alpha_\BS \bD_\BS(\gamma_b)\bF_\BS  \big(\bI_\BS(\gamma_b) \odot(\bb_\BS(\tau_b) (\zz(\thetab)\odot \bc_\BS(\gamma_b))^\trp )\big) + \nonumber \\ & \sqrt{P_\SAT}\alpha_\SAT \bD_\SAT(\gamma_s) \bF_{\SAT}(\gamma_\SAT)\big(\bI_\SAT(\gamma_s)\odot (\bb_\SAT(\tau_s^\res) \bc_\SAT^\trp( \gamma_s))\big) + \bN, \notag
\end{align}
where $\bD_\BS(\gamma_b) = \bD^q_\BS(\gamma_b, \epsilon_b = 0)$, $\bD_\SAT(\gamma_s) = \bD^q_\SAT(\gamma_s, \epsilon_s = 0)$ are
\begin{align}
[\bD_\BS(\gamma_b)]_{k,k} = &e^{-j2\pi f_c \gamma_b {T_0}k/{N}}, \\
 [\bD_\SAT(\gamma_s)]_{k,k} = & e^{-j2\pi f_c(1- ({1-\gamma_s})/({1-\bar \psi_\bs}) ){T_0} k/{N} },
\end{align}
while  $\bF_\BS = \bF^q_\BS(\gamma_\BS, \epsilon_\BS = 0) \approx \bF^q_\BS(\gamma_\BS = 0, \epsilon_\BS = 0)$, $\bF_\SAT(\gamma_\SAT) = \bF_\SAT^q(\gamma_\SAT, \epsilon_\SAT=0)$ are\footnote{Since $\gamma_\BS$ is much smaller than $\gamma_\SAT$, and Doppler precompensation in our setup affects only the carrier frequency, we can approximate $\bF^q_\BS(\gamma_\BS, \epsilon_\BS = 0) \approx \bF^q_\BS(\gamma_\BS = 0, \epsilon_\BS = 0) = \bF_\BS$, effectively reducing this matrix to the IDFT matrix. If Doppler precompensation was applied at both the carrier frequency and the subcarrier levels, $\bF_\SAT^q(\gamma_\SAT, \epsilon_\SAT=0)$ would similarly reduce to the IDFT matrix.}
\begin{align}
   [\bF_\BS]_{k,n} &= \frac{1}{\sqrt{N}}e^{j 2\pi nk/N}, ~\text{for even}~~ n,\\
   [\bF_\SAT(\gamma_\SAT)]_{k,n} &= \frac{1}{\sqrt{N}}e^{j 2\pi (1-\gamma_\SAT) nk/N}, ~\text{for odd}~~ n. \label{eq_FSAT}
\end{align}
Finally, $\bI_\BS(\gamma_b) = \bI^q_\BS(\gamma_b, \epsilon_b = 0)$, $\bI_\SAT(\gamma_s) = \bI^q_\SAT(\gamma_s, \epsilon_s=0)$ are
\begin{align}
    [\bI_{\BS, m}(\gamma_b)]_{n,m} = & e^{-j 2\pi \gamma_b mn (g+1)}, ~\text{for even}~~ n,\\
 [\bI_{\SAT, m}(\gamma_s)]_{n,m} = & e^{-j 2\pi \gamma_s mn (g+1) }, ~\text{for odd}~~ n,
\end{align}
and $\bc_\BS(\gamma_b) = \bc_\BS^q(\gamma_b, \epsilon_b=0)$, $\bc_\SAT(\gamma_s) = \bc_\SAT^q(\gamma_s, \epsilon_s=0)$ are
\begin{align}
    [\bc_\BS(\gamma_b)]_m  &=  e^{-j2\pi f_c \gamma_b m \Ts}, \label{eq_c_BS}\\
    [\bc_\SAT(\gamma_s)]_m & = e^{-j2\pi f_c((1-(1-\gamma_s)/(1-\bar\psi_\bs))m\Ts)}, \label{eq_c_SAT}
\end{align}
and $[\bz(\thetab)]_m = \bw'^\trp(t'=m\Ts + \Tcp)\aab(\thetab)$.

%\subsubsection{Model C}
\subsubsection{Model without ICI (CCFODnoICI)}
As the next step towards simplifying \eqref{eq_TM}, we neglect ICI and Doppler effect on subcarriers over fast time (related to \eqref{eq_FSAT}) as well. Therefore the received signal can be modeled as
\begin{align} \label{eq_model_c}
     \bY & = \sqrt{P_{\BS}}\alpha_\BS \bF_\BS \big(\bI_\BS(\gamma_b) \odot(\bb_\BS(\tau_b) (\zz(\thetab)\odot \bc_\BS(\gamma_b))^\trp )\big)  \nonumber \\ & +\sqrt{P_\SAT}\alpha_\SAT \bF_\SAT\big(\bI_\SAT(\gamma_s)\odot (\bb_\SAT(\tau_s^\res) \bc_\SAT^\trp(\gamma_s))\big) + \bN.
\end{align}
Here, $\bF_\SAT = \bF_\SAT^q(\gamma_s = 0, \epsilon_s = 0).$
%\subsubsection{Model B}
\subsubsection{Model with only Slow Doppler (SlowD)}
To further simplify the received signal, we neglect the intersubcarrier Doppler effect as well, leading to 
\begin{align} \label{eq_model_b}
     \bY & = \sqrt{P_\BS}\alpha_\BS \bF_\BS (\bb_\BS(\tau_b) (\zz(\thetab)\odot \bc_\BS(\gamma_b))^\trp )  \nonumber \\ & +\sqrt{P_\SAT}\alpha_\SAT \bF_\SAT (\bb_\SAT(\tau_s^\res) \bc_\SAT^\trp(\gamma_s)) + \bN,
\end{align}
in which only the slow-time Doppler effect is present, meaning that the phase changes for every OFDM symbol and is constant throughout $N$ subcarriers in each OFDM symbol. This model is common in the OFDM \ac{ISAC} literature \cite{koivunen2024multicarrier}.
%\subsubsection{Model A}
\subsubsection{Communication Coherence Interval Model (Comm)}
Finally, we derive the simplest model, where mobility is assumed but the CFO and Doppler effects are considered negligible due to their minimal impact over the short observation period. In this model, these effects are treated as a fixed phase over the entire frame, which can be absorbed into the channel gain. Consequently, neither slow nor fast time Doppler nor CFO effects are accounted for, and no phase rotation over time is observed. Under these assumptions, the received signal can be expressed as follows:
\begin{align} \label{eq_model_a}
    \bY = &\sqrt{P_\BS}\alpha_\BS \bF_\BS (\bb_\BS(\tau_b) \zz^\trp(\thetab))  + \sqrt{P_\SAT}\alpha_\SAT \bF_\SAT (\bb_\SAT(\tau_s^\res) \mathbf{1}_M^\trp) + \bN.
\end{align}
This model is common in the communication literature \cite{alkhateeb2016frequency} and also the positioning literature under low mobility \cite{wymeersch2022radio}.

% -------
\section{Estimation Algorithms} \label{sec_methods}
In this section, we present the maximum likelihood (ML) estimation algorithms corresponding to the proposed simplified models in Section \ref{subsec_simpmodels}, starting with the simplest model and progressing to more complex ones.
\subsection{Channel Parameter Estimation}
\subsubsection{Model Comm}
The channel-domain parameter vector is
\begin{align} \label{eq_chi_modela}
    \Chib_\ch^a = [\alphareBS, \alphaimBS, \alphareSAT, \alphaimSAT, \tau_b, \tau_s^\res, \thetab^\trp]^\trp \in \realsetone{8}.
\end{align}
Since the satellite-UE and BS-UE paths do not share any common unknown parameters, each path can be analyzed independently. Leveraging the structure of model Comm, along with the orthogonal subcarriers employed by the BS and the satellite, the contributions from the BS and satellite can be separated as follows:
\begin{align}\label{eq_BS-SAT-sep1}
    \tilde \bY_\BS & = \bF_\BS^\her \bY, \\
    \tilde \bY_\SAT & = \bF_\SAT^\her \bY. \label{eq_BS-SAT-sep2}
\end{align}
To estimate $\tau_s^\res$, the following problem needs to be solved: 
\begin{align} \label{eq_modela_solution_SAT}
    [\hat\tau_s^\res, \hat\alpha_\SAT] = \argmin_{\tau, \alpha_\SAT} \norm{\tilde \bY_\SAT -   \alpha_\SAT\sqrt{P_\SAT}\bb_\SAT(\tau)\mathbf{1}_M^\trp}^2,
\end{align}
which results in
\begin{align} \label{eq_modela_solution_SAT}
    \hat\tau_s^\res = \argmin_\tau \norm{\tilde \bY_\SAT - \hat\alpha_\SAT(\tau)\sqrt{P_\SAT}\bb_\SAT(\tau)\mathbf{1}_M^\trp}^2,
\end{align}
where
\begin{align}
    \hat{\alpha}_\SAT(\tau) = \frac{\norm{\tilde\bY_\SAT \mathbf{1}_M \bb_\SAT^\her(\tau) }^2}{\sqrt{P_s}\norm{\bb_\SAT(\tau)\mathbf{1}_M^\trp}^2} = \frac{\norm{\tilde\bY_\SAT \mathbf{1}_M \bb_\SAT^\her(\tau) }^2}{\sqrt{P_s}MN/2}. 
\end{align}
It follows with \eqref{eq_modela_solution_SAT} that
\begin{align} \label{eq_est_taus_modela}
    \hat\tau_s^\res = \argmax_{\tau} \abs{\bb_\SAT^\her(\tau) \tilde\bY_\SAT \mathbf{1}_M}.
\end{align}
The problem is initially solved by performing a 1D grid search, which can then be refined using a 1D quasi-Newton algorithm to obtain more accurate estimates of $\tau_s^\res$.

%We can refine our estimate using \eqref{eq_modela_solution_SAT}. 
To estimate $\tau_b$ and $\thetab$,  the following problem needs to be solved:
\begin{align} \label{eq_modela_solution_BS}
    [\hat\tau_b, \hat\thetab^\trp, \hat\alpha_\BS] = \argmin_{\tau, \thetab, \alpha_\BS} \norm{\tilde \bY_\BS - \alpha_\BS\sqrt{P_\BS}\bb_\BS(\tau)\bz^\trp(\thetab)}^2.
    %\nonumber \\ &\norm{\tilde \bY_\BS - \alpha_\BS(\tau, \thetab)\sqrt{P_\BS}\bb_\BS(\tau)\bz^\trp(\thetab)}^2.
\end{align}
which results in
\begin{align}
     &[\hat\tau_b, \hat\thetab^\trp] = \argmin_{ \tau, \thetab}  \norm{\tilde \bY_\BS - \hat\alpha_\BS(\tau, \thetab)\sqrt{P_\BS}\bb_\BS(\tau) \zz^\trp(\thetab)}^2.
\end{align}
where
\begin{align}
  \hat\alpha_\BS(\tau, \thetab) = \frac{\norm{\tilde\bY_\BS(\zz(\thetab)^\conj\bb_\BS^\her(\tau)) }^2}{\sqrt{P_\BS}\norm{\bb_\BS(\tau)\zz^\trp(\thetab)}^2}.
\end{align}
The estimation problem \eqref{eq_modela_solution_BS} can be solved using a separate 1D and 2D grid search, where the initial search % employs non-coherent integration over the \ac{OFDM} symbols (slow-time)
 is to estimate the delay
\begin{align} \label{eq_est_taub_modela}
    \hat\tau_b = \argmax_{\tau} \abs{\bb_\BS^\her(\tau) \tilde \bY_\BS \mathbf{1}_M},
\end{align}
followed by 2D AoD search 
% coherent integration over the sub-carriers to estimate the 2D \acp{AoD}
\begin{align}
    \hat\thetab = \argmax_{\thetab}\abs{\bb_\BS^\her(\hat\tau_b)\tilde \bY_\BS \bz^\conj(\thetab)}.
\end{align}
We can refine our estimates using \eqref{eq_modela_solution_BS} by performing the quasi-Newton algorithm. 

\subsubsection{Model SlowD} In model SlowD, the channel-domain parameter vector is
\begin{align}\label{eq_chi_modelb}
    \Chib_\ch^b = [{\Chib_\ch^a}^\trp, \gamma_b, \gamma_s]^\trp \in \realsetone{10}.
\end{align}
Similar to model Comm, the BS-UE and satellite-UE paths can be separated due to the use of orthogonal subcarriers using \eqref{eq_BS-SAT-sep1} and \eqref{eq_BS-SAT-sep2}. To estimate $\tau_s^\res$ and $\gamma_s$, the following problem needs to be solved:
\begin{align} \label{eq_modelb_solution_SAT}
    &[\hat\tau_s^\res, \hat\gamma_s] = \nonumber \\ &\argmin_{\tau, \gamma} \norm{\tilde \bY_\SAT - \hat\alpha_\SAT(\tau, \gamma)\sqrt{P_\SAT}\bb_\SAT(\tau)\bc_\SAT(\gamma)^\trp}^2,
\end{align}
where 
\begin{align}
    \hat\alpha_s(\tau, \gamma) = \frac{\norm{\tilde\bY_\SAT(\bc_\SAT^\conj(\gamma) \bb_\SAT^\her(\tau)) }^2}{\sqrt{P_s}\norm{\bb_\SAT(\tau)\bc_\SAT(\gamma)^\trp}^2} .
\end{align}
Since $\tau_s^\res$ is related to the subcarriers and $\gamma_s$ is associated with the slow-time samples, these parameters can be estimated independently using separate grid searches. To estimate $\tau_s^\res$, we %can perform non-coherent integration over slow-time and 
utilize \eqref{eq_est_taus_modela}. % to determine $\hat\tau_s^\res$.
Subsequently, this result can be used to %perform coherent integration over the sub-carriers and 
estimate $\gamma_s$ accordingly:
%\begin{align}
%    \hat\gamma_s = \argmin_\gamma \norm{\bb^\her_\SAT(\hat\tau_s^\res)\tilde \bY_\SAT - \hat\alpha_\SAT(\hat\tau_s^\res, \gamma)\sqrt{P_\SAT}\bc_\SAT^\trp(\gamma)}^2
%\end{align}
%which results in
\begin{align} \label{eq_est_gammas_modelb}
    \hat\gamma_s = \argmax_{\gamma} \abs{\bb_\SAT^\her(\hat\tau_s^\res) \tilde \bY_\SAT\bc_\SAT^\conj(\gamma)}.
\end{align}

For estimating parameters in the BS-UE path, the following problem needs to be solved:
\begin{align} \label{eq_modelb_solution_BS}
    &[\hat\tau_b, \gamma_b, \hat\thetab^\trp] = \argmin_{ \tau, \gamma, \thetab} \nonumber \\ &  \norm{\tilde \bY_\BS - \hat\alpha_\BS(\tau, \gamma, \thetab) \sqrt{P_\BS}(\bb_\BS(\tau) (\zz(\thetab)\odot \bc_\BS(\gamma))^\trp )}^2,
\end{align}
%which results in
%\begin{align}
%     &[\hat\tau_b, \gamma_b, \hat\thetab^\trp] = \argmin_{ \tau, \gamma, \thetab} \nonumber \\ & \norm{\tilde \bY_\BS - \alpha_\BS(\tau, \gamma, \thetab)\sqrt{P_\BS}(\bb_\BS(\tau_b) (\zz(\thetab)\odot \bc_\BS(\gamma_b))^\trp )}^2.
%\end{align}
where
\begin{align}
    \hat\alpha_\BS(\tau, \gamma, \thetab) = \frac{\norm{\tilde\bY_\BS((\zz^\conj(\thetab)\odot \bc^\conj_\BS(\gamma))\bb_\BS^\her(\tau)) }^2}{\sqrt{P_\BS}\norm{\bb_\BS(\tau)(\zz(\thetab)\odot \bc_\BS(\gamma))^\trp}^2}.
\end{align}
It can be observed that while $\tau_b$ is associated with subcarrier dimension, both $\thetab$ and $\gamma_b$ vary over slow-time domain. This variation leads to angle-Doppler coupling in the slow-time domain, and to tackle the estimation problem, a 1D grid search followed by a 3D grid search (2D angle + Doppler) is required in the basic approach. The 1D grid search is used to estimate $\tau_b$, and the subsequent 3D grid search jointly estimates $\thetab$ and $\gamma_b$. 
This complexity can be reduced by designing the beamforming matrix with repetitive angles over a subset of the observations, such as the first $P$ samples, similar to the solution in \cite{ercan2024ris}. For this subset, the beamforming matrix remains fixed, meaning that no angle information is embedded in these observations. As a result, Doppler can be estimated independently of the angles. Once $\gamma_b$ is estimated from this subset, the full set of observations can then be used to estimate the angles $\thetab$, effectively decoupling the 3D grid search into a 1D Doppler search and a 2D angle search. 
Therefore, %through non-coherent integration over slow-time and a 1D delay search, 
an estimate for $\tau_b$ is obtained using \eqref{eq_est_taub_modela}. 
% For estimating parameters in the BS-UE path, it is important to note the presence of angle-Doppler coupling in the slow-time domain. One approach to the estimation process can begin by using a 1D grid search with non-coherent integration to estimate $\tau_b$ Following this, a 3D grid search (2D angle + Doppler) can be performed to obtain coarse estimates of $\thetab$ and $\gamma_b$. To simplify this search, the beamforming matrix can be designed with repetitive angles over a portion of the observations, such as the first $P$ samples. This approach enables the estimation of $\gamma_b$ independently of the angles. Once $\gamma_b$ has been estimated, the angle estimation can be refined using the entire set of observations, \cite{ercan2024ris}. This strategy effectively reduces the complexity of the 3D grid search by converting it into a 1D search for Doppler and a 2D search for angles.
% \begin{align} \label{eq_est_taub_modelb}
%    \hat\tau_b = \argmax_{\tau} \abs{\bb_\BS^\her(\tau) \bF^\her_{\BS}\bY \mathbf{1}_M},
% \end{align}
Furthermore, an estimate of $\gamma_b$ can be found by %coherently integrating over the sub-carriers, using the estimated delay, and 
leveraging the first $P$ slow-time samples of $\tilde\bY_\BS$:
\begin{align} \label{eq_est_gammab_modelb}
    \hat\gamma_b = \argmax_{\gamma} \abs{\bb_\BS^\her(\hat\tau_b) [\tilde \bY_\BS]_{(:, 1:P)}[\bc_\BS^\conj(\gamma)]_{(1:P)}},
\end{align}
and then the full set of observations can be utilized %with coherent detection 
to estimate $\thetab$ as 
\begin{align} \label{eq_est_theta_modelb}
    \hat\thetab = \argmax_{\thetab}\abs{\bb_\BS^\her(\hat\tau_b)\tilde \bY_\BS(\bc_\BS^\conj(\hat\gamma_b)\odot\bz^\conj(\thetab))}.
\end{align}
Finally, by performing the quasi-Newton algorithm we can refine our estimates.

\subsubsection{Model CCFODnoICI} The channel-domain parameter vector in model CCFODnoICI is identical to that in model SlowD, as $\Chib_\ch^b $. The process begins by separating the BS-UE and satellite-UE contributions, using \eqref{eq_BS-SAT-sep1} and \eqref{eq_BS-SAT-sep2}.
To estimate $\tau_s^\res$ and $\gamma_s$, the following problem needs to be solved:
\begin{align} \label{eq_modelc_solution_SAT}
    &[\hat\tau_s^\res, \hat\gamma_s] = \argmin_{\tau, \gamma} \nonumber \\ & \norm{\tilde \bY_\SAT - \hat\alpha_\SAT(\tau, \gamma)\sqrt{P_\SAT}\bI_\SAT(\gamma) \odot (\bb_\SAT(\tau)\bc_\SAT(\gamma)^\trp)}^2,
\end{align}
where
\begin{align}
      \hat\alpha_s(\tau, \gamma) = \frac{\norm{\tilde\bY_\SAT(\bI_\SAT^\her(\gamma)\odot\bc_\SAT^\conj(\gamma) \bb_\SAT^\her(\tau)) }^2}{\sqrt{P_s}\norm{\bI_\SAT^\her(\gamma)\odot(\bb_\SAT(\tau)\bc_\SAT(\gamma)^\trp)}^2}.
\end{align}
 The primary distinction between model CCFODnoICI and model SlowD lies in the inclusion of the intersubcarrier Doppler effect ($\bI_\BS(\gamma_b)$ and $\bI_\SAT(\gamma_s)$). Given that a significant portion of $\gamma_s$ is already known from $\bar\psi_\bs$, this prior knowledge can be leveraged to partially mitigate the intersubcarrier Doppler effect in the satellite-UE contribution. This can facilitate satellite-UE parameter estimation. We can estimate $\gamma_s$% using non-coherent integration 
 as 
\begin{align} \label{eq_est_gammas_modelc}
    \hat\gamma_s = \argmax_{\gamma} \abs{\mathbf{1}^\trp_{N/2}  (\tilde\bY_\SAT \odot \bI^\conj_\SAT(\bar\psi_\bs) )\bc_\SAT^\conj(\gamma)}
\end{align}
and estimate $\tau_s^\res$% using coherent integration 
 as follows
\begin{align} \label{eq_est_taus_modelc}
    \hat\tau_s^\res = \argmax_{\tau} \abs{\bb_\SAT^\her(\tau) (\tilde\bY_\SAT \odot \bI^\conj_\SAT(\hat\gamma_s))\bc_\SAT^\conj(\hat\gamma_s)}.
\end{align}
The term $ \bI_\SAT(\gamma_s) $, initially approximated using a proxy to simplify the grid search, is then fully incorporated during the refinement step to improve the estimation accuracy.
% The term $\bI_\SAT(\gamma_s)$ is then utilized to refine these  estimates.

To estimate $\tau_b$, $\gamma_b$ and $\thetab$, the following problem needs to be solved:
\begin{align} \label{eq_modelc_solution_BS}
    &[\hat\tau_b, \gamma_b, \hat\thetab^\trp] = \argmin_{\tau, \gamma, \thetab} 
 ||\tilde \bY_\BS - \nonumber \\ & \hat\alpha_\BS(\tau, \gamma, \thetab)\sqrt{P_\BS}\bI_\BS(\gamma) \odot(\bb_\BS(\tau) (\zz(\thetab)\odot \bc_\BS(\gamma))^\trp )||^2,
\end{align}
where
\begin{align}
    \hat\alpha_\BS(\tau, \gamma, \thetab) = \frac{\norm{\tilde\bY_\BS(\bI_\BS^\her(\gamma)\odot(\zz^\conj(\thetab)\odot \bc^\conj_\BS(\gamma))\bb_\BS^\her(\tau)) }^2}{\sqrt{P_\BS}\norm{\bI_\BS(\gamma)\odot(\bb_\BS(\tau)(\zz(\thetab)\odot \bc_\BS(\gamma))^\trp)}^2}.
\end{align}
Coarse estimates are first obtained by neglecting the term $ \bI_\BS(\gamma_b) $ to simplify the initial grid search, using \eqref{eq_est_taub_modela}, \eqref{eq_est_gammab_modelb}, and \eqref{eq_est_theta_modelb}. These estimates are then refined by reintroducing $ \bI_\BS(\gamma_b) $ and solving the full objective in \eqref{eq_modelc_solution_BS}.
%Coarse estimates can be obtained by initially neglecting the presence of $\bI_\BS(\gamma_b)$ and applying \eqref{eq_est_taub_modela}, \eqref{eq_est_gammab_modelb} and \eqref{eq_est_theta_modelb}. These estimates can then be refined using \eqref{eq_modelc_solution_BS}.
\subsubsection{Model CCFOD} The channel-domain parameter vector in model CCFOD is the same as the one in model SlowD and model CCFODnoICI. The key difference between model CCFOD and other simplified models is that due to the existence of ICI represented by $\bD_\BS(\gamma_b)$ and $\bD_\SAT(\gamma_s)$, there may be leakage between the BS-UE and satellite-UE paths. As a result, the two paths can only be partially separated using simple processing as outlined in \eqref{eq_BS-SAT-sep1} and \eqref{eq_BS-SAT-sep2}. To estimate $\gamma_b$, $\tau_b$ and $\thetab$, we need to solve 
\begin{align} \label{eq_modeld_solution_BS}
    [\hat\tau_b, \gamma_b, &\hat\thetab^\trp] = \argmin_{\tau, \gamma, \thetab} 
 ||\bF_\BS^\her\bD_\BS^\her(\gamma) \bY -  \\ &  \hat\alpha_\BS(\tau, \gamma, \thetab)\sqrt{P_\BS}\bI_\BS(\gamma) \odot(\bb_\BS(\tau) (\zz(\thetab)\odot \bc_\BS(\gamma))^\trp )||^2,\nonumber
\end{align}
where
\begin{align}
   \hat\alpha_\BS(\tau, \gamma, \thetab) = \frac{\norm{\tilde\bY_\BS(\bI_\BS^\her(\gamma)\odot(\zz^\conj(\thetab)\odot \bc^\conj_\BS(\gamma))\bb_\BS^\her(\tau)) }^2}{\sqrt{P_\BS}\norm{\bI_\BS(\gamma)\odot(\bb_\BS(\tau)(\zz(\thetab)\odot \bc_\BS(\gamma))^\trp)}^2}.
\end{align}
To estimate $\gamma_s$ and $\tau_s^\res$ we proceed by
\begin{align} \label{eq_modeld_solution_SAT}
    [\hat\tau_s^\res, \hat\gamma_s] = &\argmin_{\tau, \gamma} ||\bF_\SAT^\her\bD_\SAT^\her(\gamma) \bY -  \nonumber \\ & \hat\alpha_\SAT(\tau, \gamma)\sqrt{P_\SAT}\bI_\SAT(\gamma) \odot (\bb_\SAT(\tau)\bc_\SAT(\gamma)^\trp)||^2,
\end{align}
where
\begin{align}
      \hat\alpha_s(\tau, \gamma) = \frac{\norm{\tilde\bY_\SAT(\bI_\SAT^\her(\gamma)\odot\bc_\SAT^\conj(\gamma) \bb_\SAT^\her(\tau)) }^2}{\sqrt{P_s}\norm{\bI_\SAT^\her(\gamma)\odot(\bb_\SAT(\tau)\bc_\SAT(\gamma)^\trp)}^2}.
      \end{align}
To tackle \eqref{eq_modeld_solution_BS}, the presence of $\bD_\BS(\gamma_b)$ and $\bI_\BS(\gamma_b)$ is temporarily neglected. Therefore, we apply \eqref{eq_BS-SAT-sep1} and \eqref{eq_BS-SAT-sep2} to partially separate two paths, and then leverage \eqref{eq_est_taub_modela}, \eqref{eq_est_gammab_modelb} and \eqref{eq_est_theta_modelb} similar to the procedures model SlowD and model CCFODnoICI to estimate $\gamma_b$, $\tau_b$ and $\thetab$. These estimates are then refined by incorporating $\bD_\BS(\gamma_b)$ and $\bI_\BS(\gamma_b)$ using \eqref{eq_modeld_solution_BS}, enabling the full reconstruction of the BS path $\tilde \bY_\BS$.

In the next step, we remove the reconstructed BS path from $\bY$ to facilitate separating the satellite-UE contribution. Now to estimate the satellite-UE related parameters ($\tau_s^\res$ and $\gamma_s$), similar to the estimation algorithm in model CCFODnoICI, we use $\bar\psi_\bs$ to eliminate the effect of $\bD_\SAT$ and $\bI_\SAT$ partly, then we estimate $\gamma_s$ and $\tau_s^\res$ as 
\begin{align} \label{eq_est_gammas_modeld}
    &\hat\gamma_s= \argmax_{\gamma} \abs{\mathbf{1}^\trp_{N/2}(\bF^\her_\SAT \bD^\her_\SAT(\bar\psi_\bs)(\bY - \tilde\bY_\BS)\odot \bI^\conj_\SAT(\bar\psi_\bs) )\bc_\SAT^\conj(\gamma)},
\end{align}
\begin{align} \label{eq_est_taus_modeld}
     & \hat\tau_s^\res  = \argmax_{\tau}  \abs{\bb_\SAT^\her(\tau) (\bF^\her_\SAT \bD^\her_\SAT(\hat\gamma_s)(\bY - \tilde\bY_\BS) \odot \bI^\conj_\SAT(\hat\gamma_s))\bc_\SAT^\conj(\hat\gamma_s)}.
\end{align}

% then we implement \eqref{eq_est_gammas_modelb} and \eqref{eq_est_taus_modela} to find coarse estimates of $\gamma_s$ and $\tau_s^\res$ and then we use $\bD_\SAT$ and $\bI_\SAT$ to find the refined estimates of the aformentioned parameters.
 \begin{remark} All the estimators presented in this section can be improved by considering the first two peaks in the first grid search for each path (e.g., \eqref{eq_est_taus_modela} and \eqref{eq_est_taub_modela} in model Comm) to account for potential leakage from the other path when using data from the true model.
\end{remark}
\begin{remark}
According to \eqref{eq_c_SAT}, %and the estimation algorithms, where the slow-time terms are primarily utilized to estimate Dopplers, 
the maximum unambiguous interval for estimating $\gamma_s$ is $(1-\bar\psi_\bs)/(f_c \Ts)$. However, due to the velocities of LEO satellites and the Earth's rotation considered in $\gamma_s$, this value may exceed the maximum range in certain scenarios.
To address this issue, the known parameter $\bar\psi_\bs$ is used to determine the integer part corresponding to the unambiguous range within the actual Doppler $\gamma_s$. Given that the UE's velocity is significantly smaller than the LEO satellite and Earth's velocities, and that the UE is in close proximity to the BS, the integer factors for $\bar\psi_\bs$ and $\gamma_s$ are expected to be identical. Consequently, only the residual Doppler needs to be estimated, after which the known integer factor can be applied to accurately retrieve $\gamma_s$.
\end{remark}

\subsection{Location, Clock Offset, Velocity and CFO Estimation}

%In model-a,  the phase does not change over the entire frame, therefore $\gamma_b, \gamma_s$ and as a result, $\eta$ and $\norm{\vv}$ are considered to be zero. 

%In this case, 
All models provide two TOA measurements and a single AOD tuple, which can be used to position the UE and estimate its clock offset \cite{Keykhosravi2020_SisoRIS}. Models SlowD, CCFODnoICI, and CCFOD, however, also provide two additional Doppler measurements, which can be utilized to estimate the UE's radial velocity and CFO.
% In all models, the estimators provide $\pp_0$ and $\Delta_{t,0}$. 
%The positional parameters are as follows:
%\begin{align} \label{eq_chi_pos_modela}
%\Chib_\pos^a =  [\alphareBS, \alphaimBS, \alphareSAT, %\alphaimSAT, \pp_0^\trp, \Delta_{t,0}]^\trp \in \realsetone{8}.
%\end{align}
%which are related via
%\begin{align} \label{eq_gammab_rel}
%    & \tau_b = \frac{\norm{\pp_0 - \ppbs}}{c} + \Delta_{t,0}, \\
%    & \tau_s^\res = \frac{\norm{\pp_0 - \ppsato}}{c} - \tau_\bs + \Delta_{t,0}. \label{eq_gammas_rel}
%\end{align}

As for estimating $\pp_0$ and $\Delta_{t,0}$, it is possible to neglect the factors $(1-\eta)(1-\psi_\BS)$ and $(1-\eta)(1-\psi_\SAT)$ in \eqref{eq_tau_b_1st} and \eqref{eq_tau_s_1st} due to their small nominal values for simplicity.% and estimate $\hat\pp_0$ and $\hat\Delta_{t,0}$ % using \eqref{eq_sim_pos_modela} and \eqref{eq_sim_deltat0_modela}
\footnote{A typical value for $\eta$ is $1$ppm, and in case of the UE moving with $80$ kph and satellite elevation angle being $\pi/4$, $\abs{\psi_\BS}$ and $\abs{\psi_\SAT}$ are in the order of $10^{-6}$ and $10^{-5}$ respectively.} Moreover, we can write the expression for the line passing the BS with AoD $\hat \thetab$ according to
%\begin{align}
 % $   \pp_0(\beta) = \ppbs + \beta {\bk(\hat\thetab)}/{\norm{\bk(\hat\thetab)}}$.
  $   \pp_0(\beta) = \ppbs + \beta \hat\uu$, where $\hat\uu = \uu(\hat\thetab)$.
%\end{align}
Therefore we can find $\hat{\pp}_0$ using
%\begin{align}
$    \hat{\pp}_0  = \pp_0(\hat\beta)$
%\end{align}
where
\begin{align} \label{eq_sim_pos_modela}
    \hat\beta &= \argmin_\beta \nonumber \\ & \bigg|{ \frac{\norm{\pp_0(\beta)-\ppbs}}{c} -  \frac{\norm{\pp_0(\beta)-\ppsato}}{c} - (\hat\tau_b - \hat\tau_s^\res - \bar\tau_\bs)}\bigg|,    
\end{align}
and $\hat{\Delta}_{t,0}$ can be found as a weighted combination of residual delays, where each residual represents the difference between the measured delay and the expected geometric delay from the estimated position:
\begin{align}
\hat{\Delta}_{t,0} & = \frac{1}{D + 1} \left( \hat{\tau}_b - \frac{\| \pp_b - \hat{\pp_0} \|}{c} \right)
\nonumber \\ & + \frac{D}{D + 1} \left( \hat{\tau}_s^{\mathrm{res}} + \bar{\tau}_{bs} - \frac{\| \pp_s - \hat{\pp}_0 \|}{c} \right),
\label{eq:clock_bias_est}
\end{align}
where $D = {\| \hat{\pp_0}- \pp_b \|}/{\| \hat{\pp_0} - \pp_s \|}$. This heuristic weighted averaging approach puts more trust in the delay estimate from the transmitter that is closer to the user, which in our case, is the BS.
% \begin{align} \label{eq_sim_deltat0_modela}
%     \hat{\Delta}_{t,0} & = \Big(\hat\tau_b - \frac{\norm{\ppbs - \hat{\pp}_0}}{c}\Big) + \Big(\hat\tau_s^\res + \bar\tau_\bs - \frac{\norm{\ppsato - \hat{\pp}_0}}{c}\Big).
% \end{align}

%In models SlowD, CCFODnoICI, and CCFOD, the Doppler $\gamma_b$ and $\gamma_s$ are estimated, so that \eqref{eq_gammab_rel} and \eqref{eq_gammas_rel} are modified to%, therefore the positional parameter vector changes to the below:
%\begin{align} \label{eq_chi_pos_modelb}
%\Chib^k_\pos =  [{\Chib^a_\pos}^\trp,\norm{\vv}, \eta]^\trp \in %\realsetone{10}, ~~ k = b,c,d,
%\end{align}
 %and below relations hold
%\begin{align}
%    & \tau_b = \frac{\norm{\pp - \ppbs}}{c} + \Delta_{t,0}(1-\eta)(1-\psi_\BS), \\
 %   & \tau_s^\res = \frac{\norm{\pp - \ppsato}}{c} - \tau_\bs + \Delta_{t,0}(1-\eta)(1-\psi_\SAT).
%\end{align}

%
As for estimating $\norm{\vv}$ and $\eta$, due to the structure of model Comm with neglecting any phase change in the transmission of $M$ OFDM symbols, it is not possible to estimate $\norm{\vv}$ and $\eta$. But in models SlowD, CCFODnoICI and CCFOD, it holds that
%\begin{align}
    $\gamma_b = \eta+(1-\eta)\psi_\BS$ and $
    \gamma_s = \eta+(1-\eta)\psi_\SAT,$
%\end{align}
where
\begin{align}
    &\psi_\BS = \frac{\norm{\vv}(\pp_0 - \ppbs)^\trp\vec{\vv}}{c\norm{\pp_0-\ppbs}}, \\
    &\psi_\SAT =  \frac{\norm{\vv}(\pp_0 - \ppsato)^\trp\vec{\vv}}{c\norm{\pp_0 - \ppsato}} - \frac{(\pp_0 - \ppsato)^\trp (\vv_\LEO - \vv_\E) }{c\norm{\pp_0 - \ppsato}},
\end{align}
according to \eqref{eq_tau_BS_SAT_quad} %\eqref{eq_vv_su}, 
and \eqref{eq_gamma_b} where $\vec{\vv} = \vv/\norm{\vv}$% and \eqref{eq_gamma_s}
. Then using estimated $\hat\pp_0$, $\hat\gamma_\BS$ and $\gamma_\SAT$, we can find $\hat{\norm{\vv}}$ and $\hat\eta$ as 
\begin{align}
& \hat{\eta} = \frac{\hat \gamma_s - ( (\hat \psi_{N,\SAT}/\hat \psi_{N,\BS})\hat\gamma_b - \tilde\psi_\SAT) }{1-((\hat \psi_{N,\SAT}/\hat \psi_{N,\BS})-\tilde\psi_\SAT)}, \\
& \hat{\norm{\vv}} = \frac{\hat \gamma_b - \hat\eta}{(1-\hat\eta)\hat\psi_{N,\BS}},
\end{align}
where
\begin{align}
    &\tilde\psi_\SAT = \frac{(\hat\pp_0 - \ppsato)^\trp (\vv_\LEO - \vv_\E) }{c\norm{\hat\pp_0 - \ppsato}},\\
    &\hat \psi_{N,\SAT} = \frac{(\hat\pp_0 - \ppsato)^\trp\vec{\vv}}{c\norm{\hat\pp_0 - \ppsato}}, \label{eq_psi_NSAT}\\
    &\hat \psi_{N,\BS} = \frac{(\hat\pp_0 - \ppbs)^\trp\vec{\vv}}{c\norm{\hat\pp_0-\ppbs}}. \label{eq_psi_NBS}
\end{align}
Here, the vector $\vec{\vv}$ denotes the known \ac{UE}'s heading and the subscript $N$ in \eqref{eq_psi_NSAT} and \eqref{eq_psi_NBS} stands for normalized, with respect to the \ac{UE}'s speed. These estimates can be further improved using gradient descent applied to the relevant cost functions.
\subsection{Complexity Analysis}
In this subsection, the proposed simplified estimators are compared in terms of computational complexity. Assuming a fixed number of grids $G$ in each dimension, and $I$ iterations for the quasi-Newton algorithm, the channel parameter estimation for the model Comm has the lowest complexity at $\mathcal{O}(GNM + G^2M + I(N+M))$. Here, the first term corresponds to the complexity of estimating the BS and satellite delays; the second term represents the complexity of 2D AoD estimation and the final term accounts for the refinement cost.
The estimator for the model SlowD has a complexity of $\mathcal{O}(GM + GP + GNM + G^2M + I(N+M))$ where the terms respectively represent the complexity of satellite Doppler estimation, BS Doppler estimation, BS and satellite delay estimation, AoD estimation, and refinement.
For the model CCFODnoICI, the complexity is given as $\mathcal{O}(N^2 + GM + GP + GNM + G^2M + I(N+M))$. The first term reflects the additional cost of considering intersubcarrier Doppler effect term in satellite Doppler estimation, while the remaining terms are analogous to those in the model SlowD.
Finally, model CCFOD has the highest complexity at $\mathcal{O}(N^2 + GM + NM + GP + GNM + G^2M + IN^2)$ Here, the third term represents the complexity of reconstructing the BS path, while the other terms are similar to the model CCFODnoICI, with the refinement step contributing $\mathcal{O}(IN^2)$, due to the inclusion of ICI in this model.

%%%%%%%%%%%%%%%%%%%%%%%%%% simulation Results %%%%%%%%%%%%%%%%%%%%%%%%%%%%%
\section{Simulation Results}\label{sec_sim}
In this section, we illustrate the performance of our estimators based on different models considering the data generated from the generative model in \eqref{eq_TM}. The goal is to analyze how accurate the simplified models are in different scenarios.

\subsection{Simulation Setup and Theoretical Bounds}
The system parameters are given in Table \ref{table_S}. The time-varying channel gains $\alpha_\BS(t)$ and $\alpha_\SAT(t)$ are modeled based on \ac{FSPL} and given by 
%\begin{align}
    $\alpha_\BS(t) = \sqrt{\cos^q(\thetab_\el(t))}{\lambda}/{(4\pi  \norm{\pp(t) - \ppbs})}$,
%\end{align}
where $q = 0.57$ \cite[Sec. 9.7.3]{stutzman2012antenna} and
%\begin{align}
        $\alpha_\SAT(t) = {\lambda}/{(4\pi \norm{\pp(t) - \ppsat(t)})}$.
%\end{align}
\begin{table}%[h] 
    \centering
    \caption{Simulations Parameters}
    \begin{tabular}{l c c}
        \hline
        \textbf{Parameter} & \textbf{Symbol} & \textbf{Value} \\
        \hline 
        Carrier frequency & $f_c$ & 2 GHz\\
        Speed of light & $c$ & $3\times 10^8$ m/s \\
        Number of subcarriers & $N$ & 3300\\
        Subcarrier spacing & $\Delta_f$ & 30 kHz \\
        Total symbol duration & $\Ts$ &  35.7 $\mu$s \\
        Symbol duration & $T_0$ & 33.3 $\mu$s\\
        Cyclic prefix duration & $\Tcp$& 2.3 $\mu$s \\
        Number of symbols & $M$ & 64 \\
        Number of antennas & $L$ & 64 \\
        LEO satellite velocity  & $\norm{\vv_\SAT}$&  7800 m/s\\
        Number of beamformer phase repetition & $P$ & 4 \\
        LEO satellite altitude & $h$ &  600 km\\
        % LEO satellite initial angle & $[\thetael, \thetaaz]$ & $[ 45^\circ, 90^\circ]$ \\
        % LET satellite initial Angle $\thetab = [\thetaaz, \thetael]^\trp$ & $[\pi/2, \pi/4]^\trp$ \\
        Earth rotation velocity& $\norm{\vv_\E}$ & 465 m/s\\
        Earth radius & $R$ & 6371 km \\
        BS position &  $\ppbs$ & $[0,0,5]^\trp$ m \\
        UE initial position & $\pp_0$ & $[20,50,1.5]^\trp$ m \\
        UE heading & $\vec{\vv}$ & $[1,0,0]^\trp$ m/s\\
        % Max UE velocity & $v = \norm{\vv}$ & 80 kph \\ 
        % UE velocity & $\vv$ & $[33, 0, 0]^\trp$ m/s \\
        % UE clock-oscillator drift & $\eta$ & $1$~ppm \\
        % UE clock-oscillator initial time bias & $\Delta_{t,0}$ & $1$~ns \\
        \hline
    \end{tabular}
    \hspace{0.5cm}
    \label{table_S}
    \end{table}
    
We employ two theoretical bounds to evaluate estimation accuracy. The first is the \ac{CRB}, which represents the minimum achievable variance of an unbiased estimator when the estimation model perfectly matches the true data generation model. However, since the estimators in our scenario are not designed based on the generative model, achieving the CRB is not guaranteed in the presence of significant model mismatch. To address this, we utilize the \ac{MCRB} as an alternative theoretical bound \cite{Fortunati2017, Ricmond2015MCRB}. The MCRB provides a lower bound on the variance of estimators under model mismatch and incorporates the effect of estimator bias. 
The CRB provides a reasonable bound for our mismatched model in the low to medium SNR regime, where estimation errors are mostly due to noise rather than model mismatches. At high SNR, however, the MCRB becomes more relevant, as the bias term dominates the bound while the variance term approaches zero, making it the appropriate theoretical bound in this regime. The adoption of these bounds in our problem is elaborated in Appendix \ref{app_CRB_MCRB}.

\subsection{Results and Discussion}
\subsubsection{Mismatched Estimation Performance}

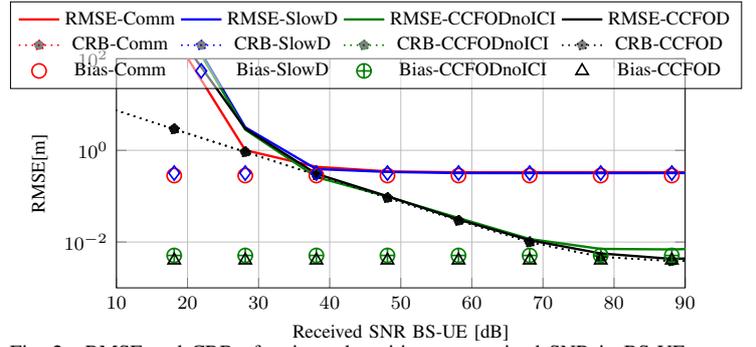
\begin{figure}
\centering
\begin{tikzpicture}
[scale=1\columnwidth/10cm,font=\footnotesize]
\begin{axis}[
    width=10cm,
    height=5cm,
    ymode=log,
    %log ticks with fixed point,
    xmin=10,
    xmax=90,
    ymin=0.001,
    ymax=100,
    xlabel={Received SNR BS-UE [dB]},
    ylabel={RMSE[m]},  
    grid=major,   
    %legend cell align=left
    legend style={at={(1.1,1.25)},anchor=north east, scale = 0.95, legend columns = 4},
    fill opacity=0.5,
    draw opacity=1,   
    text opacity=1
]

\addplot[draw=red,line width=1pt] %mark repeat=2]
table[x=SNR, y=RMSE, col sep=comma]{tikzfig/data/pos_model_a_vsPBS.txt};
\addlegendentry{RMSE-Comm}

\addplot[draw=blue, line width=1pt] %mark repeat=2]
table[x=SNR, y=RMSE, col sep=comma]{tikzfig/data/pos_model_b_vsPBS.txt};
\addlegendentry{RMSE-SlowD}

\addplot[draw=green!50!black, line width=1pt] %mark repeat=2]
table[x=SNR, y=RMSE, col sep=comma]{tikzfig/data/pos_model_c_vsPBS.txt};
\addlegendentry{RMSE-CCFODnoICI}

\addplot[draw=black, line width=1pt] %mark repeat=2]
table[x=SNR, y=RMSE, col sep=comma]{tikzfig/data/pos_model_d_vsPBS.txt};
\addlegendentry{RMSE-CCFOD}

\addplot[draw=red, dotted, mark = *, line width=0.7pt]%, mark repeat=2]
table[x=SNR, y=PEB, col sep=comma]{tikzfig/data/pos_model_a_vsPBS.txt};
\addlegendentry{CRB-Comm}

\addplot[draw=blue, dotted, mark = *, line width=0.7pt]%, mark repeat=2]
table[x=SNR, y=PEB, col sep=comma]{tikzfig/data/pos_model_b_vsPBS.txt};
\addlegendentry{CRB-SlowD}

\addplot[draw=green!50!black, dotted, mark = *, line width=0.7pt]%, mark repeat=2]
table[x=SNR, y=PEB, col sep=comma]{tikzfig/data/pos_model_c_vsPBS.txt};
\addlegendentry{CRB-CCFODnoICI}

\addplot[draw=black, dotted, mark = *, line width=0.7pt]%, mark repeat=2]
table[x=SNR, y=PEB, col sep=comma]{tikzfig/data/pos_model_d_vsPBS.txt};
\addlegendentry{CRB-CCFOD}

\addplot[only marks, draw=red, dashed, mark = o, mark size=3, mark options={solid}, line width=0.7pt]%, mark repeat=2]
table[x=SNR, y=Bias, col sep=comma]{tikzfig/data/pos_model_a_vsPBS.txt};
\addlegendentry{Bias-Comm}

\addplot[only marks, draw=blue, mark = diamond, mark options={solid}, mark size=3, line width=0.7pt]%, mark repeat=2]
table[x=SNR, y=Bias, col sep=comma]{tikzfig/data/pos_model_b_vsPBS.txt};
\addlegendentry{Bias-SlowD}

\addplot[only marks, draw=green!50!black, mark = oplus, dashed, mark options={solid}, mark size=3, line width=0.7pt]%, mark repeat=2]
table[x=SNR, y=Bias, col sep=comma]{tikzfig/data/pos_model_c_vsPBS.txt};
\addlegendentry{Bias-CCFODnoICI}

\addplot[only marks, draw=black, mark options={solid}, mark size=3, mark = triangle, line width=0.7pt]%, mark repeat=2]
table[x=SNR, y=Bias, col sep=comma]{tikzfig/data/pos_model_d_vsPBS.txt};
\addlegendentry{Bias-CCFOD}

\end{axis}
\end{tikzpicture}

%\end{document}
\vspace{-3mm}
\caption{RMSE and CRB of estimated position vs. received SNR in BS-UE path. {Since the legends in all figures are similar to that of Fig. \ref{fig_pos_vs_PBS}, they are omitted in the subsequent figures for clarity. Bias values are included where relevant and omitted otherwise.}% for model Comm-CCFOD with data from true model.
}%\vspace{-5mm}
\label{fig_pos_vs_PBS}
\end{figure}

\begin{figure}
\centering
\begin{tikzpicture}
[scale=1\columnwidth/10cm,font=\footnotesize]
\begin{axis}[
    width=10cm,
    height=5cm,
    ymode=log,
    %log ticks with fixed point,
    xmin=10,
    xmax=90,
    ymin=0.001,
    ymax=100,
    xlabel={Received SNR satellite-UE [dB]},
    ylabel={RMSE [m]},  
    grid=major,   
    %legend cell align=left
    legend style={at={(1.1,1.25)},anchor=north east, scale = 0.95, legend columns = 4},
    fill opacity=0.5,
    draw opacity=1,   
    text opacity=1
]

\addplot[draw=red,line width=1pt] %mark repeat=2]
table[x=SNR, y=RMSE, col sep=comma]{tikzfig/data/pos_model_a_vsPSAT.txt};
%\addlegendentry{RMSE-Comm}

\addplot[draw=blue, line width=1pt] %mark repeat=2]
table[x=SNR, y=RMSE, col sep=comma]{tikzfig/data/pos_model_b_vsPSAT.txt};
%\addlegendentry{RMSE-SlowD}

\addplot[draw=green!50!black, line width=1pt] %mark repeat=2]
table[x=SNR, y=RMSE, col sep=comma]{tikzfig/data/pos_model_c_vsPSAT.txt};
%\addlegendentry{RMSE-CCFODnoICI}

\addplot[draw=black, line width=1pt] %mark repeat=2]
table[x=SNR, y=RMSE, col sep=comma]{tikzfig/data/pos_model_d_vsPSAT.txt};
%\addlegendentry{RMSE-CCFOD}

\addplot[draw=red, dotted, mark = *,line width=0.7pt]%, mark repeat=2]
table[x=SNR, y=PEB, col sep=comma]{tikzfig/data/pos_model_a_vsPSAT.txt};
%\addlegendentry{CRB-Comm}

\addplot[draw=blue, dotted, mark = *,line width=0.7pt]%, mark repeat=2]
table[x=SNR, y=PEB, col sep=comma]{tikzfig/data/pos_model_b_vsPSAT.txt};
%\addlegendentry{CRB-SlowD}

\addplot[draw=green!50!black, mark = *,dotted, line width=0.7pt]%, mark repeat=2]
table[x=SNR, y=PEB, col sep=comma]{tikzfig/data/pos_model_c_vsPSAT.txt};
%\addlegendentry{CRB-CCFODnoICI}

\addplot[draw=black, dotted, mark = *,line width=0.7pt]%, mark repeat=2]
table[x=SNR, y=PEB, col sep=comma]{tikzfig/data/pos_model_d_vsPSAT.txt};
%\addlegendentry{CRB-CCFOD}

\addplot[only marks, draw=red, dotted, mark = o, mark size=3, mark options={solid}, line width=0.7pt]%, mark repeat=2]
table[x=SNR, y=Bias, col sep=comma]{tikzfig/data/pos_model_a_vsPSAT.txt};
%\addlegendentry{Bias-Comm}

\addplot[only marks, draw=blue, mark = diamond, mark options={solid}, mark size=3,dotted, line width=0.7pt]%, mark repeat=2]
table[x=SNR, y=Bias, col sep=comma]{tikzfig/data/pos_model_b_vsPSAT.txt};
%\addlegendentry{Bias-SlowD}

\addplot[only marks, draw=green!50!black, mark = oplus, dotted, mark options={solid}, mark size=3, line width=0.7pt]%, mark repeat=2]
table[x=SNR, y=Bias, col sep=comma]{tikzfig/data/pos_model_c_vsPSAT.txt};
%\addlegendentry{Bias-CCFODnoICI}

\addplot[only marks, draw=black, dotted, mark options={solid}, mark size=3, mark = triangle, dotted, line width=0.7pt]%, mark repeat=2]
table[x=SNR, y=Bias, col sep=comma]{tikzfig/data/pos_model_d_vsPSAT.txt};
%\addlegendentry{Bias-CCFOD}

\end{axis}
\end{tikzpicture}

%\end{document}
\vspace{-2mm}
\caption{RMSE and CRB of estimated position vs. received SNR in the satellite-UE path.% for model Comm-CCFOD with data from the true model.
}%\vspace{-5mm}
\label{fig_pos_vs_PSAT}
\end{figure}
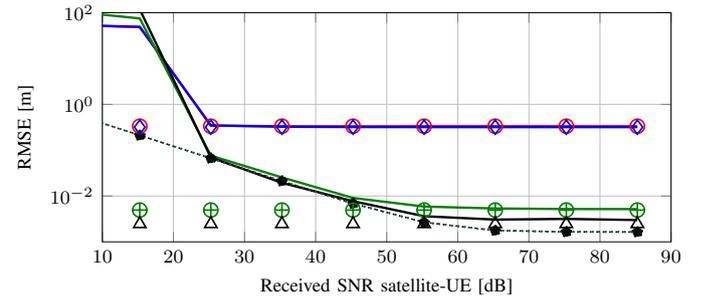

First, we analyze the \ac{RMSE}, \ac{CRB} and bias terms of estimated UE's position for all simplified models, considering data generated from the true model versus received \ac{SNR}. To better understand the effect of BS and satellite transmit power, two figures are presented to illustrate the trends: Fig. \ref{fig_pos_vs_PBS} shows the case where the satellite transmit power is fixed at $P_\SAT = 65$ dBm, while Fig. \ref{fig_pos_vs_PSAT} depicts the case where the BS transmit power is fixed at $P_\BS = 35$ dBm. In Fig. \ref{fig_pos_vs_PBS} the BS transmit power changes from $-40$ dBm to $50$ dBm, and in Fig. \ref{fig_pos_vs_PSAT} the satellite transmit power changes from $10$ dBm to $80$ dBm. In both figures, the velocity magnitude is set to $\norm{\vv} = 15$ kph, the CFO is $\eta = 10^{-8}$ and the satellite is located at an elevation angle of $\thetab^\SAT_\el = 88^{\circ}$.
It is observed that the CRB for all four models coincides exactly in both figures. The reason is that the combined Doppler and CFO ($\gamma_b$ and $\gamma_s$) contains marginal position information compared to the other channel-domain parameters $\tau_s$, $\tau_b$ and $\thetab$. The performance of the Comm and SlowD models is very similar, indicating that while accounting for slow-time Doppler is expected to improve AoD estimation and consequently positioning, its impact is overshadowed by the absence of inter-subcarrier Doppler compensation in this scenario. This observation is further reinforced by the performance improvement seen in the CCFODnoICI model compared to Comm and SlowD. The performance of models CCFODnoICI and CCFOD approaches the CRB at high SNR, showing importance of compensating for inter-subcarrier Doppler effect. In contrast, a significant gap is observed between the bias of models Comm and SlowD and the CRB at high SNR, highlighting that the performance of these two models is limited by model mismatch. Notably, except for high SNR, the performance of the CCFODnoICI and CCFOD models is similar, indicating that accounting for ICI has a negligible impact in this scenario at lower SNRs. However, the difference becomes more noticeable at high SNR.

The analyses presented above apply to the specific scenario described. In the following subsections, different scenarios are explored to provide a deeper understanding of the behavior of our algorithm.
%Among the models, model Comm and model SlowD exhibit the largest gap, with model CCFODnoICI outperforming them and model CCFOD achieves the best performance, as expected. 

\begin{comment}
    
\begin{figure}
\centering
\input{tikzfig/datacode_jan20th/pos_vs_PSAT}\vspace{-2mm}
\caption{RMSE and CRB of estimated position vs. received SNR in satellite-UE path for model Comm-CCFOD with data from true model - satellite West to East, $\pp_0 = [0, 50, 1.5]^\trp$.}%\vspace{-5mm}
\label{fig_pos_vs_PSAT_jan20th}
\end{figure}
\end{comment}

\begin{comment}
\begin{figure}
\centering
\input{tikzfig/datacode_jan20th/pos_vs_PBS}\vspace{-3mm}
\caption{RMSE and CRB of estimated position vs. received SNR in BS-UE path for model Comm-CCFOD with data from true model - satellite West to East, $\pp_0 = [0, 50, 1.5]^\trp$.}%\vspace{-5mm}
\label{fig_pos_vs_PBS_jan20th}
\end{figure}
\end{comment}

\subsubsection{Impact of CFO}\label{subsec:CFO}
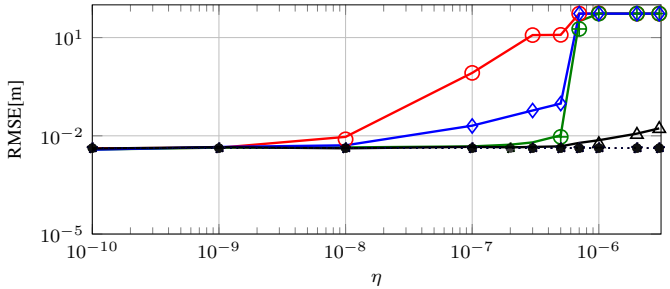
\begin{figure}
\centering
\begin{tikzpicture}
[scale=1\columnwidth/10cm,font=\footnotesize]
\begin{comment}

%\begin{axis}[
%    width=10cm,
%    height=7cm,
%    ymode=log,
    xmode=log,
    log ticks with fixed point,
    log basis x=10,
    xmin=0,
    xmax=4e-6,
    ymin=0.001,
    scaled x ticks=base 10:-6, % Factor out 1e-6
    ymax=100,
    xlabel={$\eta$},
    xtick scale label code/.code={$\times 10^{-6}$}, % Add the scale label near the x-axis arrow
    ylabel={[m]},  
    grid=major,   
    %legend cell align=left
    legend style={at={(1,1)},anchor=north east, scale = 0.95, legend columns = 2},
    fill opacity=0.5,
    draw opacity=1,   
    text opacity=1
]
\end{comment}

\begin{loglogaxis}[
        width = 10cm,
        height = 5cm,
        log basis x=10,
        log basis y=10,
        % For the log scale, pick valid positive ranges:
        xmin=1e-10,    xmax=3.1e-6,
        ymin=1e-5, ymax=1e2,
        xlabel={$\eta$},
        ylabel={RMSE[m]},
        % Factor out 10^-6 from the x-axis labels:
        scaled x ticks = base 10:-6,
        % Place the '× 10^-6' near the x-axis arrow:
        xtick scale label code/.code={$\times 10^{-6}$},
        % Optional: control the spacing or positions of ticks
        % xtick = {1e-8,1e-7,1e-6,1e-5},
        % ytick = {1e-4,1e-2,1,1e2},
        % axis line styles:
        % axis x line=bottom,
        % axis y line=left,
        grid=major,   
        legend style={at={(0.785,1.13)},anchor=north east, scale = 0.95, legend columns = 2},
       fill opacity=0.5,
       draw opacity=1,   
       text opacity=1
    ]
    
\addplot[draw=red,line width=1pt] %mark repeat=2]
table[x=eta, y=RMSE, col sep=comma]{tikzfig/data/pos_model_a_vseta.txt};
%\addlegendentry{RMSE-model a}

\addplot[draw=green!50!black, line width=1pt] %mark repeat=2]
table[x=eta, y=RMSE, col sep=comma]{tikzfig/data/pos_model_c_vseta.txt};
%\addlegendentry{RMSE-model c}

\addplot[only marks, draw=red, dotted, mark = o, mark size=3, mark options={solid}, line width=0.7pt]%, mark repeat=2]
table[x=eta, y=Bias, col sep=comma]{tikzfig/data/pos_model_a_vseta.txt};
%\addlegendentry{Bias-model a}

\addplot[only marks, draw=green!50!black, mark = oplus, dotted, mark options={solid}, mark size=3, line width=0.7pt]%, mark repeat=2]
table[x=eta, y=Bias, col sep=comma]{tikzfig/data/pos_model_c_vseta.txt};
%\addlegendentry{Bias-model c}

\addplot[draw=red, mark = *,dotted, line width=0.7pt]%, mark repeat=2]
table[x=eta, y=PEB, col sep=comma]{tikzfig/data/pos_model_a_vseta.txt};
%\addlegendentry{CRB-model a}

\addplot[draw=green!50!black, mark = *,dotted, line width=0.7pt]%, mark repeat=2]
table[x=eta, y=PEB, col sep=comma]{tikzfig/data/pos_model_c_vseta.txt};
%\addlegendentry{CRB-model c}

\addplot[draw=blue, line width=1pt] %mark repeat=2]
table[x=eta, y=RMSE, col sep=comma]{tikzfig/data/pos_model_b_vseta.txt};
%\addlegendentry{RMSE-model b}

\addplot[draw=black, line width=1pt] %mark repeat=2]
table[x=eta, y=RMSE, col sep=comma]{tikzfig/data/pos_model_d_vseta.txt};
%\addlegendentry{RMSE-model d}

\addplot[only marks, draw=blue, mark = diamond, mark options={solid}, mark size=3,dotted, line width=0.7pt]%, mark repeat=2]
table[x=eta, y=Bias, col sep=comma]{tikzfig/data/pos_model_b_vseta.txt};
%\addlegendentry{Bias-model b}

\addplot[only marks, draw=black, dotted, mark options={solid}, mark size=3, mark = triangle, dotted, line width=0.7pt]%, mark repeat=2]
table[x=eta, y=Bias, col sep=comma]{tikzfig/data/pos_model_d_vseta.txt};
%\addlegendentry{Bias-model d}

\addplot[draw=blue, dotted, mark = *,line width=0.7pt]%, mark repeat=2]
table[x=eta, y=PEB, col sep=comma]{tikzfig/data/pos_model_b_vseta.txt};
%\addlegendentry{CRB-model b} 

\addplot[draw=black, dotted, mark = *,line width=0.7pt]%, mark repeat=2]
table[x=eta, y=PEB, col sep=comma]{tikzfig/data/pos_model_d_vseta.txt};
%\addlegendentry{CRB-model d}

% \end{axis}
\end{loglogaxis}
\end{tikzpicture}

%\end{document}
\vspace{-3mm}
\caption{RMSE and CRB of estimated position vs. CFO% ($\eta$)% for model Comm-CCFOD with data from true model
.}%\vspace{-3mm}
\label{fig_pos_vs_eta}
\end{figure}

In Fig. \ref{fig_pos_vs_eta}, the positioning performance of the four models is analyzed as a function of the CFO. For this evaluation, the satellite is positioned at the zenith, and the user is assumed to be stationary. The transmit powers of the BS and satellite are set to $P_\BS = 35$ dBm and $P_\SAT = 65$ dBm, respectively. The positioning CRB for all four models coincides, as expected, and all models achieve the CRB at low values of $\eta$. Therefore, the bias term is only presented for cases where the RMSE deviates from the CRB. By increasing $\eta$, model Comm, then model SlowD and model CCFODnoICI 
%and finally model CCFOD 
would introduce large estimation errors, but it is model CCFOD that takes the CFO into account in the ICI term as well as in modified Fourier matrix $\bF_\SAT(\gamma_\SAT)$, 
%without any precompensation
therefore, it is more robust towards CFO variations compared to the other models. 
The reason is that model Comm entirely ignores the presence of CFO, making it the most susceptible to CFO variations among all models. In contrast, the other models account for CFO to some extent, enabling them to tolerate higher CFO values. Among these, model CCFOD demonstrates the highest robustness. An important observation is that for higher values of CFO, the CCFOD model still achieves performance close to the CRB, resulting in centimeter-level positioning accuracy. This indicates that even for large CFO values, a complex estimation algorithm incorporating time-varying \acp{AoD} and second-order terms is not necessary. Instead, our most advanced estimation algorithm is sufficient to achieve centimeter-level positioning accuracy. In case more relaxed requirements on positioning accuracy, other simplified models could be used.
%, as its modeling includes ICI effects. %It appears that the difference in model SlowD and model CCFODnoICI in considering CFO in intersubcarrier Doppler effect term results in negligible difference in the behavior of these two models%why exactly??
%, but it is model CCFOD which takes the CFO into account in ICI term without any precompensation, therefore it is more robust towards CFO variations comparing to the other models.

\begin{comment}
\begin{figure}
\centering
\input{tikzfig/pos_vs_eta_highSNR}\vspace{-4mm}
\caption{RMSE and CRB of estimated position vs. CFO ($\eta$) for model Comm-CCFOD with data from true model at $P_\BS = 10$ dBm, $P_\SAT = 60$ dBm.}%\vspace{-5mm}
\label{fig_pos_vs_eta_highSNR}
\end{figure}
\end{comment}

\begin{comment}
\begin{figure}
\centering
\input{tikzfig/datacode_jan20th/pos_vs_eta}\vspace{-4mm}
\caption{RMSE and CRB of estimated position vs. CFO ($\eta$) for model Comm-CCFOD with data from true model  - satellite West to East, $\pp_0 = [0, 50, 1.5]^\trp$.}%\vspace{-3mm}
\label{fig_pos_vs_eta_jan20th}
\end{figure}
\end{comment}

\subsubsection{Impact of UE Speed} \label{subsec:UEV}
\begin{figure}
\centering
\begin{tikzpicture}
[scale=1\columnwidth/10cm,font=\footnotesize]
\begin{comment}

%\begin{axis}[
%    width=10cm,
%    height=7cm,
%    ymode=log,
    xmode=log,
    log ticks with fixed point,
    log basis x=10,
    xmin=0,
    xmax=4e-6,
    ymin=0.001,
    scaled x ticks=base 10:-6, % Factor out 1e-6
    ymax=100,
    xlabel={$\eta$},
    xtick scale label code/.code={$\times 10^{-6}$}, % Add the scale label near the x-axis arrow
    ylabel={[m]},  
    grid=major,   
    %legend cell align=left
    legend style={at={(1,1)},anchor=north east, scale = 0.95, legend columns = 2},
    fill opacity=0.5,
    draw opacity=1,   
    text opacity=1
]
\end{comment}

\begin{axis}[
        width = 10cm,
        height = 5cm,
        % For the log scale, pick valid positive ranges:
        ymode=log,
        xmin=0,    xmax=23,
        ymin=1e-3, ymax=0.1,
        xlabel={$\norm{\vv}$ [m/s]},
        ylabel={RMSE[m]},        
        % axis line styles:
        % axis x line=bottom,
        % axis y line=left,
        grid=major,   
        legend style={at={(1,0.9)},anchor=north east, scale = 0.95, legend columns = 1},
        fill opacity=0.5,
        draw opacity=1,   
        text opacity=1
    ]
    
\addplot[draw=red,line width=1pt] %mark repeat=2]
table[x=normV, y=RMSE, col sep=comma]{tikzfig/data/pos_model_a_vsnormV.txt};
%\addlegendentry{RMSE-model a}

\addplot[draw=red, dashed, mark = *,,line width=0.7pt]%, mark repeat=2]
table[x=normV, y=PEB, col sep=comma]{tikzfig/data/pos_model_a_vsnormV.txt};
%\addlegendentry{CRB-model a}

\addplot[draw=blue, line width=1pt] %mark repeat=2]
table[x=normV, y=RMSE, col sep=comma]{tikzfig/data/pos_model_b_vsnormV.txt};
%\addlegendentry{RMSE-model b}

\addplot[draw=blue, dashed,mark = *, line width=0.7pt]%, mark repeat=2]
table[x=normV, y=PEB, col sep=comma]{tikzfig/data/pos_model_b_vsnormV.txt};
%\addlegendentry{CRB-model b} 

\addplot[draw=green!50!black, line width=1pt] %mark repeat=2]
table[x=normV, y=RMSE, col sep=comma]{tikzfig/data/pos_model_c_vsnormV.txt};
%\addlegendentry{RMSE-model c}

\addplot[draw=green!50!black, mark = *,dashed, line width=0.7pt]%, mark repeat=2]
table[x=normV, y=PEB, col sep=comma]{tikzfig/data/pos_model_c_vsnormV.txt};
%\addlegendentry{CRB-model c}

\addplot[only marks, draw=red, mark = o, mark size=3, mark options={solid}, line width=0.7pt]%, mark repeat=2]
table[x=normV, y=Bias, col sep=comma]{tikzfig/data/pos_model_a_vsnormV.txt};
%\addlegendentry{Bias-model a}

\addplot[only marks, draw=green!50!black, mark = oplus, mark options={solid}, mark size=3, line width=0.7pt]%, mark repeat=2]
table[x=normV, y=Bias, col sep=comma]{tikzfig/data/pos_model_c_vsnormV.txt};
%\addlegendentry{Bias-model c}

\addplot[draw=black, line width=1pt] %mark repeat=2]
table[x=normV, y=RMSE, col sep=comma]{tikzfig/data/pos_model_d_vsnormV.txt};
%\addlegendentry{RMSE-model d}

\addplot[draw=black, mark = *,dashed, line width=0.7pt]%, mark repeat=2]
table[x=normV, y=PEB, col sep=comma]{tikzfig/data/pos_model_d_vsnormV.txt};
%\addlegendentry{CRB-model d}

\addplot[only marks, draw=blue, mark = diamond, mark options={solid}, mark size=3, line width=0.7pt]%, mark repeat=2]
table[x=normV, y=Bias, col sep=comma]{tikzfig/data/pos_model_b_vsnormV.txt};
%\addlegendentry{Bias-model b}

\addplot[only marks, draw=black, mark options={solid}, mark size=3, mark = triangle, dashed, line width=0.7pt]%, mark repeat=2]
table[x=normV, y=Bias, col sep=comma]{tikzfig/data/pos_model_d_vsnormV.txt};
%\addlegendentry{Bias-model d}

\end{axis}
\end{tikzpicture}

%\end{document}
\vspace{-2mm}
\caption{RMSE and CRB of estimated position vs. UE's speed $\norm{\vv}$% for model Comm-CCFOD with data from true model
.}
\label{fig_pos_vs_normV}
\end{figure}
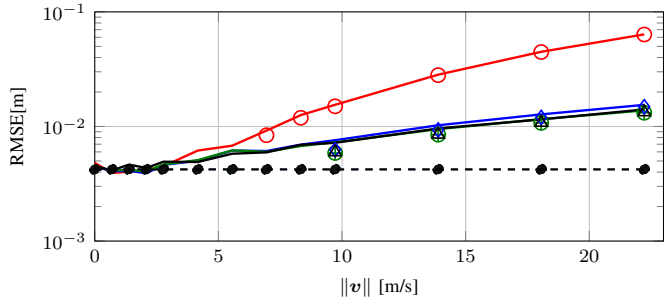

In Fig. \ref{fig_pos_vs_normV}, the behavior of our four models is analyzed with varying speeds. We consider $\eta = 0$ in this scenario and the satellite is located at the zenith with the transmit powers of the BS and satellite set to $P_\BS = 35$ dBm and $P_\SAT = 65$ dBm, respectively. Model Comm performs poorly when the velocity is as low as $3 ~\text{m/s}$, as expected. In contrast, the other three models demonstrate strong robustness to variations in velocity due to the inclusion of $\gamma_b$ and $\gamma_s$ in their models. It is important to note that the impact of CFO and radial velocity is conveyed through Doppler shifts, making their effects similar when the satellite is located at the zenith and we investigate the effect of either CFO or radial velocity individually. Specifically in such case, the positioning accuracy when the UE's speed is $15$ m/s is comparable to that when $\eta = 15/(3\times 10^{8}) \approx 0.5 \times 10^{-7}$. Therefore, for practical values of the UE's speed, since the difference between models SlowD, CCFODnoICI and CCFOD is marginal, and model SlowD is sufficient.
% Models CCFODnoICI and CCFOD demonstrate strong robustness to variations in velocity with models Comm and SlowD performing similarly.

\begin{comment}
\begin{figure}
\centering
\input{tikzfig/pos_vs_normV_highSNR}\vspace{-4mm}
\caption{RMSE and CRB of estimated position vs. magnitude of velocity ($\norm{\vv}$) for model Comm-CCFOD with data from true model at $P_\BS = 10$ dBm, $P_\SAT = 60$ dBm.}%\vspace{-5mm}
\label{fig_pos_vs_normV_highSNR}
\end{figure}
\end{comment}

\begin{comment}
\begin{figure}
\centering
\input{tikzfig/datacode_jan20th/pos_vs_normV}\vspace{-2mm}
\caption{RMSE and CRB of estimated position vs. magnitude of velocity ($\norm{\vv}$) for model Comm-CCFOD with data from true model - satellite West to East, $\pp_0 = [0, 50, 1.5]^\trp$.}\vspace{-3mm}
\label{fig_pos_vs_normV_jan20th}
\end{figure}
\end{comment}

\subsubsection{Impact of Satellite Elevation} \label{subsec:satEl}

\begin{figure}
\centering
\begin{tikzpicture}
[scale=1\columnwidth/10cm,font=\footnotesize]
\begin{comment}

%\begin{axis}[
%    width=10cm,
%    height=7cm,
%    ymode=log,
    xmode=log,
    log ticks with fixed point,
    log basis x=10,
    xmin=0,
    xmax=4e-6,
    ymin=0.001,
    scaled x ticks=base 10:-6, % Factor out 1e-6
    ymax=100,
    xlabel={$\eta$},
    xtick scale label code/.code={$\times 10^{-6}$}, % Add the scale label near the x-axis arrow
    ylabel={[m]},  
    grid=major,   
    %legend cell align=left
    legend style={at={(1,1)},anchor=north east, scale = 0.95, legend columns = 2},
    fill opacity=0.5,
    draw opacity=1,   
    text opacity=1
]
\end{comment}

\begin{axis}[
        width = 10cm,
        height = 5cm,
        % For the log scale, pick valid positive ranges:
        ymode=log,
        xmin=40,    xmax=140,
        ymin=1e-3, ymax=100,
        xlabel={$\thetab_\el^\SAT$},
        ylabel={RMSE[m]},        
        % axis line styles:
        % axis x line=bottom,
        % axis y line=left,
        grid=major,   
        legend style={at={(1.3,0.85)},anchor=north east, scale = 0.75, legend columns = 1},
        fill opacity=0.5,
        draw opacity=1,   
        text opacity=1
    ]
    
\addplot[draw=red,line width=1pt] %mark repeat=2]
table[x=theta_el, y=RMSE, col sep=comma]{tikzfig/data/pos_model_a_vsthetaSAT.txt};
%\addlegendentry{RMSE-model a}

\addplot[draw=red, dotted,mark = *, line width=0.7pt]%, mark repeat=2]
table[x=theta_el, y=PEB, col sep=comma]{tikzfig/data/pos_model_a_vsthetaSAT.txt};
%\addlegendentry{CRB-model a}

\addplot[draw=blue, line width=1pt] %mark repeat=2]
table[x=theta_el, y=RMSE, col sep=comma]{tikzfig/data/pos_model_b_vsthetaSAT.txt};
%\addlegendentry{RMSE-model b}

\addplot[draw=blue, dotted,mark = *, line width=0.7pt]%, mark repeat=2]
table[x=theta_el, y=PEB, col sep=comma]{tikzfig/data/pos_model_b_vsthetaSAT.txt};
%\addlegendentry{CRB-model b} 

\addplot[draw=green!50!black, line width=1pt] %mark repeat=2]
table[x=theta_el, y=RMSE, col sep=comma]{tikzfig/data/pos_model_c_vsthetaSAT.txt};
%\addlegendentry{RMSE-model c}

\addplot[draw=green!50!black,mark = *, dotted, line width=0.7pt]%, mark repeat=2]
table[x=theta_el, y=PEB, col sep=comma]{tikzfig/data/pos_model_c_vsthetaSAT.txt};
%\addlegendentry{CRB-model c}

\addplot[only marks, draw=red, dotted, mark = o, mark size=3, mark options={solid}, line width=0.7pt]%, mark repeat=2]
table[x=theta_el, y=Bias, col sep=comma]{tikzfig/data/pos_model_a_vsthetaSAT.txt};
%\addlegendentry{Bias-model a}

\addplot[only marks, draw=green!50!black, mark = oplus, mark options={solid}, mark size=3, line width=0.7pt]%, mark repeat=2]
table[x=theta_el, y=Bias, col sep=comma]{tikzfig/data/pos_model_c_vsthetaSAT.txt};
%\addlegendentry{Bias-model c}

\addplot[draw=black, line width=1pt] %mark repeat=2]
table[x=theta_el, y=RMSE, col sep=comma]{tikzfig/data/pos_model_d_vsthetaSAT.txt};
%\addlegendentry{RMSE-model d}

\addplot[draw=black, dotted,mark = *, line width=0.7pt]%, mark repeat=2]
table[x=theta_el, y=PEB, col sep=comma]{tikzfig/data/pos_model_d_vsthetaSAT.txt};
%\addlegendentry{CRB-model d}

\addplot[only marks, draw=blue, mark = diamond, mark options={solid}, mark size=3,dotted, line width=0.7pt]%, mark repeat=2]
table[x=theta_el, y=Bias, col sep=comma]{tikzfig/data/pos_model_b_vsthetaSAT.txt};
%\addlegendentry{Bias-model b}

 \addplot[only marks, draw=black, dotted, mark options={solid}, mark size=3, mark = triangle, dotted, line width=0.7pt]%, mark repeat=2]
table[x=theta_el, y=Bias, col sep=comma]{tikzfig/data/pos_model_d_vsthetaSAT.txt};
%\addlegendentry{Bias-model d}

\end{axis}
\end{tikzpicture}

%\end{document}
%\vspace{-4mm}
\caption{RMSE and CRB of estimated position vs. satellite elevation angle% for model Comm-CCFOD with data from true model
.}%
%\vspace{-5mm}
\label{fig_pos_vs_theta}
\end{figure}
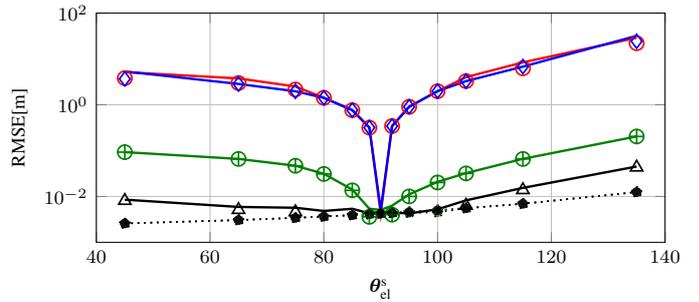

Fig. \ref{fig_pos_vs_theta} demonstrates the behavior of our models versus satellite elevation angles. The UE is assumed to be stationary, with $\eta = 0$ and the transmit powers of the BS and satellite are fixed at $P_\BS = 35$ dBm and $P_\SAT = 65$ dBm, respectively. Under these conditions, all four models perform well when $\thetab_\el^\SAT = \pi/2$. This is because $\gamma_s \approx 0$ due to the negligible radial velocity between the satellite and the UE, and $\gamma_b = 0$ as the UE is stationary. Consequently, model Comm, which does not account for Doppler effects, performs effectively. With $\gamma_b = 0$, $\gamma_s \approx 0$, $\epsilon_b = 0$, and $\epsilon_s \approx 0$, there is no mismatch between the generative model and any of the simplified models.
However, as the satellite elevation angle deviates from $\pi/2$, the performance of models Comm and SlowD deteriorates. Model Comm performs poorly because it entirely neglects $\gamma_s$, which increases rapidly as the satellite moves away from the zenith. While model SlowD accounts for $\gamma_s$, it only considers it in the slow-time domain, leaving delay estimation across subcarriers susceptible to errors. In contrast, models CCFODnoICI and CCFOD consider inter-subcarrier Doppler effects, which spans over the slow time as well as the subcarrier domains. By estimating $\gamma_s$ and compensating for these terms, models CCFODnoICI and CCFOD significantly outperform models Comm and SlowD.
%
% The similar performance of models CCFODnoICI and CCFOD can be attributed to their key distinction —the ICI term—in satellite-UE path depends on the residual Doppler $(1-(1-\gamma_s)/(1-\bar\psi_\bs))$, which is negligible when $\eta = 0$ and $\vv = 0$. The marginal difference between these two models for satellite elevation angle close to $\pi/4$ and $3\pi/4$ can be explained as follows: Model CCFOD achieves the CRB for all values of $\theta_\el^\SAT$ whereas model CCFODnoICI exhibits marginal deviation with the CRB, which is due to different ways these two model handle the separation of BS and satellite contributions. Model CCFODnoICI uses plain IDFT matrix whereas model CCFOD considers the first term in $\bF_\SAT(\gamma_\SAT)$ which is the dominant term in $\bF^q_\SAT(\gamma_\SAT, \epsilon_\SAT)$ for satellite elevation angle near $\pi/4$ and $3\pi/4$.

It is worth noting that $\gamma_s$ increases as the satellite elevation angle diverges further from the zenith. According to \eqref{eq_Fq_SAT}, a larger $\gamma_s$ causes the first term in $\bF^q_\SAT(\gamma_\SAT, \epsilon_\SAT)$ to diverge more substantially from the IDFT matrix, due to the fast-time Doppler effect. In models Comm, SlowD and CCFODnoICI, the IDFT matrix is used to separate BS-UE and satellite-UE contributions based on \eqref{eq_model_c}, \eqref{eq_model_b}, and \eqref{eq_model_a}. This divergence makes it increasingly challenging for the estimators to separate the BS-UE path from the satellite-UE path, ultimately degrading localization performance, whereas model CCFOD \eqref{eq_model_d} considers the first term in $\bF_\SAT(\gamma_\SAT)$ which is the dominant term in $\bF^q_\SAT(\gamma_\SAT, \epsilon_\SAT)$ for satellite elevation angle away from zenith.

\begin{comment}
    
\begin{figure}
\centering
\input{tikzfig/pos_vs_theta_highSNR}%\vspace{-4mm}
\caption{RMSE and CRB of estimated position vs. satellite elevation angle for model Comm-CCFOD with data from true model at $P_\BS = 10$ dBm, $P_\SAT = 60$ dBm}%
%\vspace{-5mm}
\label{fig_pos_vs_theta_highSNR}
\end{figure}
\end{comment}

\begin{comment}
\begin{figure}
\centering
\input{tikzfig/datacode_jan20th/pos_vs_theta}%\vspace{-4mm}
\caption{RMSE and CRB of estimated position vs. satellite elevation angle for model Comm-CCFOD with data from true model at $P_\BS = 10$ dBm, $P_\SAT = 60$ dBm - satellite West to East, $\pp_0 = [0, 50, 1.5]^\trp$.}%
%\vspace{-5mm}
\label{fig_pos_vs_theta_jan20th}
\end{figure}
\end{comment}
\begin{remark}
   In general, the choice of the appropriate estimation algorithm depends on factors such as the required positioning accuracy, the expected scenario, and the complexity preferences. Sec. \ref{subsec:CFO} to \ref{subsec:satEl} provide a general overview of how different models perform in various scenarios, and the results can serve as a guideline for model selection.
\end{remark}
\subsubsection{Estimation Performance of Other Parameters}
\begin{figure}%[h]
\centering
\begin{tikzpicture}
[scale=1\columnwidth/10cm,font=\footnotesize]
\begin{axis}[
    width=10cm,
    height=5cm,
    ymode=log,
    xmin=10,
    xmax=90,
    ymin=0.0001,
    ymax=1000,
    xlabel={Received SNR satellite-UE [dB]},
    ylabel={RMSE[m/sec]},  
    grid=major,   
    %legend cell align=left
    legend style={at={(0,0)},anchor=south west, scale = 0.95, legend columns = 3},
    fill opacity=0.5,
    draw opacity=1,   
    text opacity=1,
    % Turn off default scientific notation
    %scaled y ticks = false,
    % Set base for log scale
    %log basis y=10,
    % Manually define your major ticks
    %ytick={1e-11,1e-10,1e-9,1e-8},
    % Provide custom labels
    %yticklabels={$10^{-11}$,$10^{-10}$,$10^{-9}$,$10^{-8}$},
    % (Optional) Add minor ticks
    %minor y tick num=4
]

%\addplot[draw=red,line width=1pt] %mark repeat=2]
%table[x=SNR, y=RMSE, col sep=comma]{tikzfig/data/normV_model_a_vsPSAT.txt};
%\addlegendentry{RMSE-model a}
\addplot[draw=blue, line width=1pt] %mark repeat=2]
table[x=SNR, y=RMSE, col sep=comma]{tikzfig/data/normV_model_b_vsPSAT.txt};
%\addlegendentry{RMSE-model b}

\addplot[only marks, draw=blue, mark = diamond, mark options={solid}, mark size=3,dotted, line width=0.7pt]%, mark repeat=2]
table[x=SNR, y=Bias, col sep=comma]{tikzfig/data/normV_model_b_vsPSAT.txt};
%\addlegendentry{Bias-model b}

\addplot[draw=blue, dotted, mark = *,line width=0.7pt]%, mark repeat=2]
table[x=SNR, y=PEB, col sep=comma]{tikzfig/data/normV_model_b_vsPSAT.txt};
%\addlegendentry{CRB-model b}

\addplot[draw=green!50!black, line width=1pt] %mark repeat=2]
table[x=SNR, y=RMSE, col sep=comma]{tikzfig/data/normV_model_c_vsPSAT.txt};
%\addlegendentry{RMSE-model c}

%\addplot[draw=red, dotted, mark = o, mark size=3, mark options={solid}, line width=0.7pt]%, mark repeat=2]
%table[x=SNR, y=Bias, col sep=comma]{tikzfig/data/normV_model_a_vsPSAT.txt};
%\addlegendentry{Bias-model a}

\addplot[only marks,draw=green!50!black, mark = oplus, dotted, mark options={solid}, mark size=3, line width=0.7pt]%, mark repeat=2]
table[x=SNR, y=Bias, col sep=comma]{tikzfig/data/normV_model_c_vsPSAT.txt};
%\addlegendentry{Bias-model c}

%\addplot[draw=red, dotted, line width=0.7pt]%, mark repeat=2]
%table[x=SNR, y=PEB, col sep=comma]{tikzfig/data/normV_model_a_vsPSAT.txt};
%\addlegendentry{CRB-model a}

\addplot[draw=green!50!black,mark = *, dotted, line width=0.7pt]%, mark repeat=2]
table[x=SNR, y=PEB, col sep=comma]{tikzfig/data/normV_model_c_vsPSAT.txt};
%\addlegendentry{CRB-model c}

\addplot[draw=black, line width=1pt] %mark repeat=2]
table[x=SNR, y=RMSE, col sep=comma]{tikzfig/data/normV_model_d_vsPSAT.txt};
%\addlegendentry{RMSE-model d}

\addplot[only marks,draw=black, dotted, mark options={solid}, mark size=3, mark = triangle, dotted, line width=0.7pt]%, mark repeat=2]
table[x=SNR, y=Bias, col sep=comma]{tikzfig/data/normV_model_d_vsPSAT.txt};
%\addlegendentry{Bias-model d}

\addplot[draw=black, dotted,mark = *, line width=0.7pt]%, mark repeat=2]
table[x=SNR, y=PEB, col sep=comma]{tikzfig/data/normV_model_d_vsPSAT.txt};
%\addlegendentry{CRB-model d}

\end{axis}
\end{tikzpicture}

%\end{document}
%\vspace{-4mm}
\caption{RMSE and CRB of estimated $\norm{\vv}$ vs. received SNR% for model SlowD, CCFODnoICI and CCFOD with data from generative model
.}%\vspace{-5mm}
\label{fig_normV_vs_PSAT}
\end{figure}
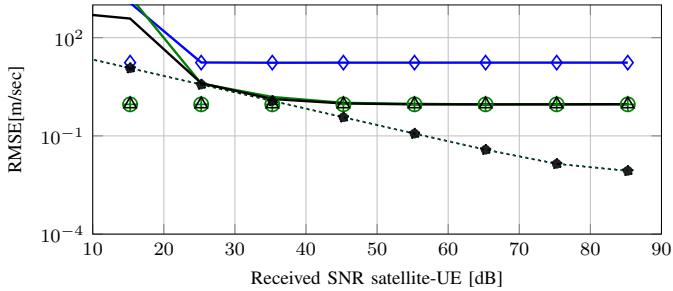

Finally, the estimation performance of $\Delta_{t,0}$, $\eta$ and $\norm{\vv}$ are illustrated in Figs. \ref{fig_delta_t0_vs_PSAT}, \ref{fig_eta_vs_PSAT}, and \ref{fig_normV_vs_PSAT}, respectively. The \ac{UE} velocity is set to $15$ kph, the BS transmit power is set to $P_\BS = 35$ dBm, with $\Delta_{t,0} = 1$ ns, $\eta = 10^{-8}$ and the satellite positioned at an elevation angle of $\thetab_\el^\SAT = 88 ^\circ$, which is the simulation scenario as the one used in Fig. \ref{fig_pos_vs_PSAT}. In Fig. \ref{fig_delta_t0_vs_PSAT}, models Comm and SlowD exhibit similar behavior, while model CCFODnoICI outperforms them, with model CCFOD almost achieving the CRB. As for estimation performance of $\eta$ and $\norm{\vv}$ in Figs. \ref{fig_eta_vs_PSAT}--\ref{fig_normV_vs_PSAT}, models CCFODnoICI and CCFOD perform similarly and outperform model SlowD as expected and model Comm is not included since it is not capable of estimating $\eta$ and $\norm{\vv}$. % All models can estimate $\Delta_{t,0}$ whereas only models $b-d$ can estimate the latter two parameters. %(\textit{there is something wrong about performance of b in the last figure, I will regenerate the results.})
\begin{figure}%[h]
\centering
\begin{tikzpicture}
[scale=1\columnwidth/10cm,font=\footnotesize]
\begin{axis}[
    width=10cm,
    height=5cm,
    ymode=log,
    % log ticks with fixed point,
    xmin=10,
    xmax=90,
    ymin=1e-14,
    ymax=1e-6,
    xlabel={Received SNR satellite-UE [dB]},
    ylabel={RMSE[sec]},  
    grid=major,   
    %legend cell align=left
    legend style={at={(1,1.1)},anchor=north east, scale = 0.95, legend columns = 2},
    fill opacity=0.5,
    draw opacity=1,   
    text opacity=1,
    % Turn off default scientific notation
    scaled y ticks = false,
    % Set base for log scale
    log basis y=10,
    % Manually define your major ticks
    % ytick={1e-14,1e-13,1e-12,1e-11,1e-10,1e-9,1e-8,1e-7},
    % Provide custom labels
   % yticklabels={$10^{-14}$,$10^{-13}$,$10^{-12}$,$10^{-11}$,$10^{-10}$,$10^{-9}$,$10^{-8}$,$10^{-7}$,$10^{-6}$},
    % (Optional) Add minor ticks
  %  minor y tick num=9
]

\addplot[draw=red,line width=1pt] %mark repeat=2]
table[x=SNR, y=RMSE, col sep=comma]{tikzfig/data/delta_t0_model_a_vsPSAT.txt};
%\addlegendentry{RMSE-model a}

\addplot[draw=green!50!black, line width=1pt] %mark repeat=2]
table[x=SNR, y=RMSE, col sep=comma]{tikzfig/data/delta_t0_model_c_vsPSAT.txt};
%\addlegendentry{RMSE-model c}

\addplot[only marks,draw=red, dotted, mark = o, mark size=3, mark options={solid}, line width=0.7pt]%, mark repeat=2]
table[x=SNR, y=Bias, col sep=comma]{tikzfig/data/delta_t0_model_a_vsPSAT.txt};
%\addlegendentry{Bias-model a}

\addplot[only marks,draw=green!50!black, mark = oplus, dotted, mark options={solid}, mark size=3, line width=0.7pt]%, mark repeat=2]
table[x=SNR, y=Bias, col sep=comma]{tikzfig/data/delta_t0_model_c_vsPSAT.txt};
%\addlegendentry{Bias-model c}

\addplot[draw=red, dotted, mark = *,line width=0.7pt]%, mark repeat=2]
table[x=SNR, y=PEB, col sep=comma]{tikzfig/data/delta_t0_model_a_vsPSAT.txt};
%\addlegendentry{CRB-model a}

\addplot[draw=green!50!black,mark = *, dotted, line width=0.7pt]%, mark repeat=2]
table[x=SNR, y=PEB, col sep=comma]{tikzfig/data/delta_t0_model_c_vsPSAT.txt};
%\addlegendentry{CRB-model c}

\addplot[draw=blue, line width=1pt] %mark repeat=2]
table[x=SNR, y=RMSE, col sep=comma]{tikzfig/data/delta_t0_model_b_vsPSAT.txt};
%\addlegendentry{RMSE-model b}

\addplot[draw=black, line width=1pt] %mark repeat=2]
table[x=SNR, y=RMSE, col sep=comma]{tikzfig/data/delta_t0_model_d_vsPSAT.txt};
%\addlegendentry{RMSE-model d}

\addplot[only marks, draw=blue, mark = diamond, mark options={solid}, mark size=3,dotted, line width=0.7pt]%, mark repeat=2]
table[x=SNR, y=Bias, col sep=comma]{tikzfig/data/delta_t0_model_b_vsPSAT.txt};
%\addlegendentry{Bias-model b}

\addplot[only marks,draw=black, dotted, mark options={solid}, mark size=3, mark = triangle, line width=0.7pt]%, mark repeat=2]
table[x=SNR, y=Bias, col sep=comma]{tikzfig/data/delta_t0_model_d_vsPSAT.txt};
%\addlegendentry{Bias-model d}

\addplot[draw=blue, dotted, mark = *,line width=0.7pt]%, mark repeat=2]
table[x=SNR, y=PEB, col sep=comma]{tikzfig/data/delta_t0_model_b_vsPSAT.txt};
%\addlegendentry{CRB-model b}

\addplot[draw=black, dotted,mark = *, line width=0.7pt]%, mark repeat=2]
table[x=SNR, y=PEB, col sep=comma]{tikzfig/data/delta_t0_model_d_vsPSAT.txt};
%\addlegendentry{CRB-model d}

\end{axis}
\end{tikzpicture}

%\end{document}
\vspace{-2mm}
\caption{RMSE and CRB of estimated $\Delta_{t,0}$ vs. received SNR% for model Comm-CCFOD with data from true model
.}%\vspace{-5mm}
\label{fig_delta_t0_vs_PSAT}
\end{figure}
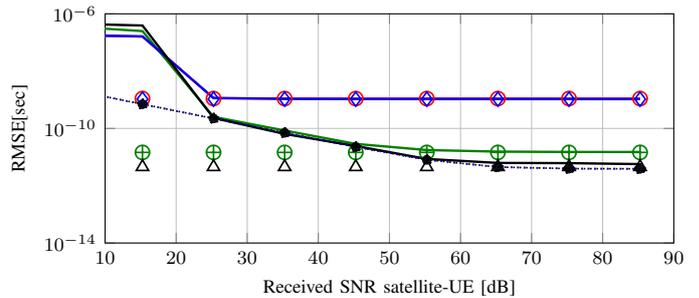

\begin{comment}
\begin{figure}
\centering
\input{tikzfig/datacode_jan20th/delta_t0_vs_PSAT}\vspace{-2mm}
\caption{RMSE and CRB of estimated $\Delta_{t,0}$ vs. received SNR for model Comm-CCFOD with data from true model - satellite West to East, $\pp_0 = [0, 50, 1.5]^\trp$.}%\vspace{-5mm}
\label{fig_delta_t0_vs_PSAT_jan20th}
\end{figure}
\end{comment}

\begin{figure}%[h]
\centering
\begin{tikzpicture}
[scale=1\columnwidth/10cm,font=\footnotesize]
\begin{axis}[
    width=10cm,
    height=5cm,
    ymode=log,
    xmin=10,
    xmax=90,
    ymin=1e-13,
    ymax=1e-5,
    xlabel={Received SNR satellite-UE [dB]},
    ylabel={RMSE},  
    grid=major,   
    %legend cell align=left
    legend style={at={(0,0)},anchor=south west, scale = 0.95, legend columns = 3},
    fill opacity=0.5,
    draw opacity=1,   
    text opacity=1,
    % Turn off default scientific notation
    scaled y ticks = false,
    % Set base for log scale
    log basis y=10,
    % Manually define your major ticks
   % ytick={1e-13,1e-12,1e-11,1e-10,1e-9,1e-8,1e-7, 1e-6,1e-5},
    % Provide custom labels
   % yticklabels={$10^{-13}$,$10^{-12}$,$10^{-11}$,$10^{-10}$,$10^{-9}$,$10^{-8}$,$10^{-7}$,$10^{-6}$,$10^{-5}$},
    % (Optional) Add minor ticks
    minor y tick num=4
]

%\addplot[draw=red,line width=1pt] %mark repeat=2]
%table[x=SNR, y=RMSE, col sep=comma]{tikzfig/data/eta_model_a_vsPSAT.txt};
%\addlegendentry{RMSE-model a}
\addplot[draw=blue, line width=1pt] %mark repeat=2]
table[x=SNR, y=RMSE, col sep=comma]{tikzfig/data/eta_model_b_vsPSAT.txt};
%\addlegendentry{RMSE-model b}

\addplot[only marks, draw=blue, mark = diamond, mark options={solid}, mark size=3,dotted, line width=0.7pt]%, mark repeat=2]
table[x=SNR, y=Bias, col sep=comma]{tikzfig/data/eta_model_b_vsPSAT.txt};
%\addlegendentry{Bias-model b}

\addplot[draw=blue, dotted,mark = *, line width=0.7pt]%, mark repeat=2]
table[x=SNR, y=PEB, col sep=comma]{tikzfig/data/eta_model_b_vsPSAT.txt};
%\addlegendentry{CRB-model b}

\addplot[draw=green!50!black, line width=1pt] %mark repeat=2]
table[x=SNR, y=RMSE, col sep=comma]{tikzfig/data/eta_model_c_vsPSAT.txt};
%\addlegendentry{RMSE-model c}

%\addplot[draw=red, dotted, mark = o, mark size=3, mark options={solid}, line width=0.7pt]%, mark repeat=2]
%table[x=SNR, y=Bias, col sep=comma]{tikzfig/data/eta_model_a_vsPSAT.txt};
%\addlegendentry{Bias-model a}

\addplot[only marks,draw=green!50!black, mark = oplus, dotted, mark options={solid}, mark size=3, line width=0.7pt]%, mark repeat=2]
table[x=SNR, y=Bias, col sep=comma]{tikzfig/data/eta_model_c_vsPSAT.txt};
%\addlegendentry{Bias-model c}

%\addplot[draw=red, dotted, line width=0.7pt]%, mark repeat=2]
%table[x=SNR, y=PEB, col sep=comma]{tikzfig/data/eta_model_a_vsPSAT.txt};
%\addlegendentry{CRB-model a}

\addplot[draw=green!50!black,mark = *, dotted, line width=0.7pt]%, mark repeat=2]
table[x=SNR, y=PEB, col sep=comma]{tikzfig/data/eta_model_c_vsPSAT.txt};
%\addlegendentry{CRB-model c}

\addplot[draw=black, line width=1pt] %mark repeat=2]
table[x=SNR, y=RMSE, col sep=comma]{tikzfig/data/eta_model_d_vsPSAT.txt};
%\addlegendentry{RMSE-model d}

\addplot[only marks,draw=black, dotted, mark options={solid}, mark size=3, mark = triangle, dotted, line width=0.7pt]%, mark repeat=2]
table[x=SNR, y=Bias, col sep=comma]{tikzfig/data/eta_model_d_vsPSAT.txt};
%\addlegendentry{Bias-model d}

\addplot[draw=black, dotted, mark = *,line width=0.7pt]%, mark repeat=2]
table[x=SNR, y=PEB, col sep=comma]{tikzfig/data/eta_model_d_vsPSAT.txt};
%\addlegendentry{CRB-model d}

\end{axis}
\end{tikzpicture}

%\end{document}
\vspace{-2mm}
\caption{RMSE and CRB of estimated $\eta$ vs. received SNR% for models SlowD, CCFODnoICI and CCFOD with data from generative model
.}%\vspace{-5mm}
\label{fig_eta_vs_PSAT}
\end{figure}
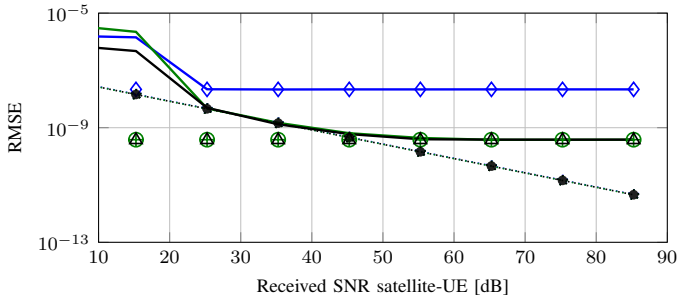

\begin{comment}
\begin{figure}
\centering
\input{tikzfig/datacode_jan20th/eta_vs_PSAT}\vspace{-2mm}
\caption{RMSE and CRB of estimated $\eta$ vs. received SNR for models SlowD, CCFODnoICI and CCFOD with data from generative model - satellite West to East, $\pp_0 = [0, 50, 1.5]^\trp$.}%\vspace{-5mm}
\label{fig_eta_vs_PSAT_jan20th}
\end{figure}
\end{comment}

By comparing Figs. \ref{fig_eta_vs_PSAT}--\ref{fig_normV_vs_PSAT} with Fig. \ref{fig_pos_vs_PSAT}, it can be observed that the second-order terms become significant at high SNR, but and only in the estimation of $\eta$ and $\norm{\vv}$. This is due to the fact that while Doppler conveys negligible positioning information, it conveys significant information for estimating $\eta$ and $\norm{\vv}$ and specifically the second-order terms (derived based on time-varying Doppler) will be crucial only when we operate at high SNR and we aim at estimating $\eta$ and $\norm{\vv}$ with very high accuracy. 
Based on Figs.~\ref{fig_eta_vs_PSAT}--\ref{fig_normV_vs_PSAT}, our CCFODnoICI and CCFOD models would achieve speed and CFO estimation accuracies of sub-$1$ m/s and sub-$0.1$ ppm respectively, therefore as long as these values meet the requirements of our application, there is no need to incorporate second-order terms.
\begin{comment}
\begin{figure}
\centering
\input{tikzfig/datacode_jan20th/normV_vs_PSAT}%\vspace{-4mm}
\caption{RMSE and CRB of estimated $\norm{\vv}$ vs. received SNR for model SlowD, CCFODnoICI and CCFOD with data from generative model - satellite West to East, $\pp_0 = [0, 50, 1.5]^\trp$.}%\vspace{-5mm}
\label{fig_normV_vs_PSAT_jan20th}
\end{figure}
\end{comment}
% Fig. 2 $\rightarrow$ RMSE, CRB and bias of position for 4 models vs. base satelite power using data from true model. 

% Fig. 3 $\rightarrow$ RMSE, CRB and bias of position for 4 models vs. base station power using data from true model. 

% Fig. 4 $\rightarrow$ RMSE, CRB (and bias) of position for 4 models vs. $\eta$ using data from true model. 

% Fig. 5 $\rightarrow$ RMSE, CRB (and bias) of position for 4 models vs. $\norm{\vv}$ using data from true model. 

% Fig. 6 $\rightarrow$ RMSE, CRB (and bias) of position for 4 models vs. $\thetael^\SAT$ using data from true model. 

% Fig. 7 $\rightarrow$ CDF of positional RMSE for 4 models in 6 different scenarios using data from true model. 

% Fig. 8 $\rightarrow$ RMSE, CRB (and bias) of $\Delta_{t,0}$ for 4 models vs. $P_\BS$ and $P_\SAT$ using data from true model. 

% Fig. 9 $\rightarrow$ RMSE, CRB (and bias) of $\norm{\vv}$ for 4 models vs. $P_\BS$ and $P_\SAT$ using data from true model. 

% Fig. 10 $\rightarrow$ RMSE, CRB (and bias) of $\eta$ for 4 models vs. $P_\BS$ and $P_\SAT$ using data from true model. 

%%%%%%%%%%%%%%%%%%%%%%%%%%%%%%%%%%%%%%%%%%%%%%%%%%%%%%%
 \section{Concluding Remarks}
This paper presented a comprehensive study on localization, velocity magnitude estimation, and synchronization for a mobile \ac{UE} in integrated cellular and non-terrestrial networks. We derived a generative model that accounts for the %slow-time, fast-time, inter-carrier interference, inter-subcarrier Doppler effects, and
time-varying Doppler and path gains effect, forming the foundation for understanding the positioning system's behavior. Building upon this, we proposed a hierarchy of four simplified models, each offering a distinct trade-off between computational complexity and performance. Estimation algorithms were developed for all the models, enabling accurate estimation of position, velocity magnitude, initial clock bias, and carrier frequency offset.

Through rigorous simulations, we demonstrated the effectiveness of the proposed models across diverse scenarios. By analyzing both the performance and complexity of each model, we can strategically select the most suitable one for different deployment conditions, optimizing processing power while maintaining high accuracy. These insights offer practical guidance for improving 6G positioning and synchronization in mobile scenarios, facilitating the seamless integration of NTNs into future wireless networks.

%%%%%%%%%%%%%%%%%%%%%%%%%%%%%%%%%%%%%%%%%%%%%%%%%%%%%%
\appendices
\section{Derivation of continuous-time generative model}\label{eq_genmodel_deriv}
Following \eqref{eq_YBS_cont}, \eqref{eq_hBS_HL} and \eqref{eq_TX_BS_BB}, $\tilde y_\BS(t)$ can be written as %following:
%\[
%\tilde y_\BS(t) = \alpha_\BS(t)\begin{bmatrix} \delta(\tau - %\tau_{\BS, 1}^p(t)) & \cdots & \delta(\tau - \tau_{\BS, L}^p(t)) \end{bmatrix} \ast
 %   \begin{bmatrix} \Re(\tilde z_{\BS,1}(t)) \\ \cdots \\ \Re(\tilde z_{\BS,L}(t)) \end{bmatrix}
 % \]  
 % where $\tilde z_{\BS,l}(t) = [\tilde \bz_{\BS}(t)]_l$. Then,
  \begin{align}
     & \tilde y_\BS(t) = \alpha_\BS(t)\Re
     \big\{\sum_{l=1}^L\tilde z_{\BS,l}(t-\tau_{\BS, l}^p(t))\big\}  \\ &=\alpha_\BS(t)\Re \big\{ \sum_{l=1}^L w_l(t-\tau_{\BS, l}^p(t))x_\BS(t-\tau_{\BS, l}^p(t))e^{j2\pi f_c (t-\tau_{\BS, l}^p(t))}\big\}.\nonumber
  \end{align}
    where $ w_{l}(t) = [ \bw(t)]_l$ and $\tilde z_{\BS,l}(t) = [\tilde \bz_{\BS}(t)]_l$.
Then,
\begin{align}
         &\tilde y_\BS(t) %=\alpha_\BS(t) \Re \bigg( e^{j2\pi f_c (t-\tau_\BS^p(t))}\nonumber \\ &\Big(\sum_{l=1}^L w_l(t-\tau_{\BS, l}^p(t))x_\BS(t-\tau_{\BS, l}^p(t))e^{-j2\pi f_c \tau_{ l}^p(t)}\Big)\bigg) \nonumber 
         \\ &
         = \Re\{ \alpha_\BS(t) \aab^\trp(\thetab(t)) \ww(t - \tau_\BS^p(t)) x_\BS\left(t - \tau_\BS^p(t)\right) e^{j 2 \pi f_c \left(t  - \tau_\BS^p(t)\right)}\} \notag . 
\end{align}
\section{Linearizing the delay propagation}\label{app_lin_tau}
We express the propagation delay as
\begin{align} \label{eq_tau_app} \tau(t) &=\frac{r(t)}{c}=\frac{\norm{\pp(t) - \pp_\Ent(t)}}{c}, \end{align}
where $\Ent \in {\BS, \SAT}$ denotes the transmitter (either the BS or the satellite). The transmitter trajectory is modeled as $\pp_\Ent(t) = \pp_\Ent + \vv_\Ent t$, with $\pp_\Ent$ and $\vv_\Ent$ representing its initial 3D position and velocity, respectively. Similarly, the receiver trajectory is given by $\pp(t) = \pp_0 + \vv t$, where $\pp_0$ and $\vv$ denote the initial 3D position and velocity of the receiver. Substituting the trajectories into \eqref{eq_tau_app}, the propagation delay can be equivalently rewritten as
\begin{align} \tau(t) &=\frac{\norm{\pp_0 + \vv_{\Ent,\UE} t - \pp_\Ent}}{c}, \end{align}
where $\vv_{\Ent,\UE}$ denotes the relative velocity between the transmitter and the receiver.
% Let's consider the expression of propagation delay as in below
% \begin{align} \label{eq_tau_app}
%         \tau(t) &=\frac{\norm{r(t)}}{c} = \frac{\norm{\pp(t) - \pp_\Ent}}{c}, % \pp_\Ent
% \end{align}
% where $\Ent \in \{\BS, \SAT\}$ and $\pp_\Ent$ denote the 3D (initial) location of the transmitter (BS or satellite). The RX trajectory can be written as $\pp(t) = \pp_0 + \vv_{\Ent,\UE} t$, where $\vv_{\Ent,\UE}$ denotes the relative velocity between the transmitter $i$ and the UE, therefore 
% \begin{align}
%     \tau(t) &=\frac{\norm{\pp_0 +\vv_{\Ent,\UE} t -\pp_\Ent }}{c}.
% \end{align}
Let's expand the nominator of \eqref{eq_tau_app}
\begin{align}
    &\norm{\pp_0 +\vv_{\Ent,\UE} t - \pp_\Ent }  = \norm{ \pp_0 - \pp_\Ent } \frac{\norm{ \pp_0 +\vv_{\Ent,\UE} t - \pp_\Ent }}{\norm{\pp_0 - \pp_\Ent }} \\ \nonumber
    & = \norm{\pp_0 - \pp_\Ent } \sqrt{\frac{({\pp_0 +\vv_{\Ent,\UE} t - \pp_\Ent })^\trp({\pp_0 +\vv_{\Ent,\UE} t - \pp_\Ent })}{\norm{\pp_0 - \pp_\Ent }^2}} \\ \label{eq_dist}
        & = \norm{\pp_0 - \pp_\Ent } \sqrt{1 + \frac{2(\pp_0 - \pp_\Ent )^\trp \vv_{\Ent,\UE}t +\norm{\vv_{\Ent,\UE} t}^2}{\norm{\pp_0 - \pp_\Ent }^2}}.
    \end{align}
Let's assume that the UE and transmitter displacement during the entire transmission block is much smaller than their initial distance, which is equivalent to the stop-and-hop assumption \cite[Ch. 2.7.2]{richards2022fundamentals}. Then, we can conclude that
\begin{align} \label{eq_smallDisplacement}
     2(\pp_0 - \pp_\Ent )^\trp \vv_{\Ent,\UE}M\Ts + \norm{\vv_{\Ent,\UE} M\Ts}^2 \ll \norm{\pp_0 - \pp_\Ent }^2.
\end{align}
We can then expand \eqref{eq_dist} using the Taylor approximation \( \sqrt{1 + x} \approx 1 + \frac{x}{2} - \frac{x^2}{8} \) for small \( x \), keeping all constant, linear-in-\( t \), and quadratic-in-\( t \) terms:
\begin{align}\label{eq_dist_approx_new}
  & \norm{\pp_0 + \vv_{\Ent,\UE} t - \pp_\Ent }  \approx \norm{\pp_0 - \pp_\Ent } \times \nonumber \\ & \left(1 + \frac{t(\pp_0 - \pp_\Ent )^\trp \vv_{\Ent,\UE}+1/2(\norm{\vv_{\Ent,\UE} t}^2 - v_{\Ent,\UE}^2 t^2)}{\norm{ \pp_0 - \pp_\Ent }^2}\right) \nonumber \\ & = \norm{\pp_0 - \pp_\Ent } +\frac{t(\pp_0 - \pp_\Ent )^\trp \vv_{\Ent,\UE} + 1/2(\norm{\vv_{\Ent,\UE} t}^2- v_{\Ent,\UE}^2t^2)}{\norm{\pp_0 - \pp_\Ent }} \nonumber \\ & =  \norm{\pp_0 - \pp_\Ent } + v_{\Ent,\UE}t +1/2 a_{\Ent,\UE} t^2 .
\end{align}
where $v_{\Ent,\UE}= (\pp_0 - \pp_\Ent )^\trp \vv_{\Ent,\UE}/\norm{\pp_0 - \pp_\Ent}$ is the initial radial velocity, and the second-order term denotes the quadratic changes in the radial distance which can be interpreted as the radial pseudo-acceleration $a_{\Ent,\UE} = (\norm{\vv_{\Ent,\UE}}^2 - v_{\Ent,\UE}^2)/\norm{\pp_0 - \pp_\Ent }$. Therefore, with the above approximation, the radial distance can be written as a constant acceleration kinematic equation, and the radial velocity and the delay will be approximated as
\begin{align}
    v_{\Ent,\UE}(t) & = v_{\Ent,\UE} + a_{\Ent,\UE} t, \nonumber \\
    \tau(t) & = r(t)/c = \tau_0 + \psi_0 t + 1/2 \mu t^2,
\end{align}
where $\tau_0 = {\norm{\pp_0 - \pp_\Ent}}/{c}$ is the initial delay, $\psi_0 = v_{\Ent,\UE}/c$ is the initial normalized Doppler shift and $\mu = a_{\Ent,\UE}/c$  is the normalized Doppler shift rate. % Moreover, Doppler $\nu(t)$ is defined as
%\begin{equation}\label{eq_Doppler}
%    \nu(t) = \frac{v_\TR(t) f_c}{c} = \psi_0 f_c + \mu{f_c}t = \nu_0 + \mu f_c t,
%\end{equation}
%where $\nu_0 = \psi_0 f_c$. With the approximation in above, Doppler will change in time. If we neglect the second-order term, Doppler will be fixed in time. Note that the above approximation only works if the time-variation of the (relative) velocity can be neglected in our time window of interest.

\section{FIM and Bias Term in MRCB} \label{app_CRB_MCRB}
The positional parameters are as follows for the model Comm
\begin{align} \label{eq_chi_pos_modela}
\Chib_\pos^a =  [\alphareBS, \alphaimBS, \alphareSAT, \alphaimSAT, \pp_0^\trp, \Delta_{t,0}]^\trp \in \realsetone{8},
\end{align} while for models  $k =$ SlowD, CCFODnoICI, and CCFOD, they are
\begin{align} \label{eq_chi_pos_modelb}
\Chib^\text{Comm}_\pos =  [{\Chib^\text{Comm}_\pos}^\trp,\norm{\vv}, \eta]^\trp \in \realsetone{10}
\end{align}

The performance bounds \ac{FIM} and bias in MCRB are detailed here. As a basic bound, we use \ac{FIM}, which is given by\cite{kay1993fundamentals}
\begin{align}
    \bF_\ch = \frac{2}{\sigma^2} \sum_{m = 0}^{M-1}\sum_{n=0}^{N-1}\Re\Big\{  \frac{\partial [\bR]_{n,m}}{\partial \chibch^k}\left( \frac{\partial [\bR]_{n,m}}{\partial \chibch^k}\right)^{\her}\Big\},
\end{align}
in which $\bR \in \complexset{N}{M}$ is the noise-free part of the received signal, $\chibch^k \in \realsetone{8}$ in case of evaluating model Comm ($k$ = Comm) and $\chibch^k \in \realsetone{10}$ \eqref{eq_chi_modela} in case of evaluating model SlowD, CCFODnoICI, and CCFOD ($k=$  SlowD, CCFODnoICI, CCFOD) \eqref{eq_chi_modelb}. We can convert $\bF_\ch$ to the positional FIM, $\bF_\pos$, corresponding to positional vector by using the Jacobian matrix
\begin{math}
    \bF_{\rm{po}} = \bJ^{\rm T}\bF_{\rm{ch}}\bJ,
\end{math}
where $\bJ$ is the Jacobian matrix with elements 
\begin{math}
    \bJ_{m,n} = {\partial [\chibch^k]_m}/{\partial[\chibch^k]_n}.
\end{math}
In case of $k=$ Comm, $\bJ \in \complexset{8}{10}$ and in case of $k=$ SlowD, CCFODnoICI, CCFOD, $\bJ \in \complexset{10}{10}$.

In case of mismatched estimation, we can find the positioning bias through MCRB for each model $k =$ Comm, SlowD, CCFODnoICI, CCFOD as
\begin{align}
     B_\pos^k & = \sqrt{\operatorname{trace}((\hat{\Chib_\pos^k}_{[5:7]})^\her \hat{\Chib_\pos^k}_{[5:7]} )},
\end{align}
where
\begin{align}
 \hat{\Chib_\pos^k} & = \argmin_{\Chib_\pos^k} [\norm{\bY_\text{NF} - \bY_{\text{NF}}^k (\Chib_\pos^k)}^2].
\end{align}
Here, $\bY_\text{NF}$ represents the noise-free received signal based on the generative model, $\bY_{\text{NF}}^k $ corresponds to the noise-free observations derived from the simplified models. The vector $\Chib_\pos^k$ is given by \eqref{eq_chi_pos_modela} and \eqref{eq_chi_pos_modelb}. For models SlowD, CCFODnoICI, and CCFOD, the estimator bias for the magnitude of the initial clock bias, velocity, and CFO can be determined as 
$ B_{\Delta_{t,0}} =|\hat{\Chib_\pos^k}_{[8]}|$, $ B_{\norm{\vv}}=|(\hat{\Chib_\pos^k}_{[9]}|$, and 
$ B_\eta =|\hat{\Chib_\pos^k}_{[10]}|$.
%\begin{align}
%      B_{\Delta_{t,0}} & = \sqrt{(\hat{\Chib_\pos^k}_{[8]})^\conj \hat{\Chib_\pos^k}_{[8]} }, \\
 %       B_{\norm{\vv}} & = \sqrt{(\hat{\Chib_\pos^k}_{[9]})^\conj \hat{\Chib_\pos^k}_{[9]} }, \\
  %        B_\eta & = \sqrt{(\hat{\Chib_\pos^k}_{[10]})^\conj \hat{\Chib_\pos^k}_{[10]} }.%
%\end{align}
% CRB works as a performance metric in case of matched estimation, meaning that the estimator models are based on the data generated model. In case of misalignment between the data generated model and the estimator model, another metric called 
\balance 
\bibliographystyle{IEEEtran}
\bibliography{IEEEabrv,Sub/ris_NTN_doppler}

\begin{thebibliography}{10}
\providecommand{\url}[1]{#1}
\csname url@samestyle\endcsname
\providecommand{\newblock}{\relax}
\providecommand{\bibinfo}[2]{#2}
\providecommand{\BIBentrySTDinterwordspacing}{\spaceskip=0pt\relax}
\providecommand{\BIBentryALTinterwordstretchfactor}{4}
\providecommand{\BIBentryALTinterwordspacing}{\spaceskip=\fontdimen2\font plus
\BIBentryALTinterwordstretchfactor\fontdimen3\font minus \fontdimen4\font\relax}
\providecommand{\BIBforeignlanguage}[2]{{%
\expandafter\ifx\csname l@#1\endcsname\relax
\typeout{** WARNING: IEEEtran.bst: No hyphenation pattern has been}%
\typeout{** loaded for the language `#1'. Using the pattern for}%
\typeout{** the default language instead.}%
\else
\language=\csname l@#1\endcsname
\fi
#2}}
\providecommand{\BIBdecl}{\relax}
\BIBdecl

\bibitem{araniti2021toward}
G.~Araniti \emph{et~al.}, ``Toward {6G} non-terrestrial networks,'' \emph{IEEE Network}, vol.~36, no.~1, pp. 113--120, 2021.

\bibitem{jiang2021road}
W.~Jiang \emph{et~al.}, ``The road towards {6G}: A comprehensive survey,'' \emph{IEEE Open Journal of the Communications Society}, vol.~2, pp. 334--366, 2021.

\bibitem{azari2022evolution}
M.~M. Azari \emph{et~al.}, ``Evolution of non-terrestrial networks from 5g to {6G}: A survey,'' \emph{IEEE communications surveys \& tutorials}, vol.~24, no.~4, pp. 2633--2672, 2022.

\bibitem{dureppagari2023ntn}
H.~K. Dureppagari \emph{et~al.}, ``{NTN}-based {6G} localization: Vision, role of {LEOs}, and open problems,'' \emph{IEEE Wireless Communications}, vol.~30, no.~6, pp. 44--51, 2023.

\bibitem{guidotti2022path}
A.~Guidotti \emph{et~al.}, ``The path to {5G}-advanced and {6G} non-terrestrial network systems,'' in \emph{2022 11th Advanced Satellite Multimedia Systems Conference and the 17th Signal Processing for Space Communications Workshop (ASMS/SPSC)}.\hskip 1em plus 0.5em minus 0.4em\relax IEEE, 2022, pp. 1--8.

\bibitem{behravan2022positioning}
A.~Behravan \emph{et~al.}, ``Positioning and sensing in {6G}: Gaps, challenges, and opportunities,'' \emph{IEEE Vehicular Technology Magazine}, vol.~18, no.~1, pp. 40--48, 2022.

\bibitem{saleh2025integrated6gtnntn}
\BIBentryALTinterwordspacing
S.~Saleh \emph{et~al.}, ``Integrated {6G TN} and {NTN} localization: Challenges, opportunities, and advancements,'' 2025. [Online]. Available: \url{https://arxiv.org/abs/2501.13488}
\BIBentrySTDinterwordspacing

\bibitem{whiton2022cellular}
R.~Whiton, ``Cellular localization for autonomous driving: A function pull approach to safety-critical wireless localization,'' \emph{IEEE Vehicular Technology Magazine}, vol.~17, no.~4, pp. 28--37, 2022.

\bibitem{del2023preliminary}
J.~A. del Peral-Rosado \emph{et~al.}, ``Preliminary field results of a dedicated {5G} positioning network for enhanced hybrid positioning,'' \emph{Engineering Proceedings}, vol.~54, no.~1, p.~6, 2023.

\bibitem{dwivedi2021positioning}
S.~Dwivedi \emph{et~al.}, ``Positioning in {5G} networks,'' \emph{IEEE Communications Magazine}, vol.~59, no.~11, pp. 38--44, 2021.

\bibitem{li2017analysis}
H.~Li \emph{et~al.}, ``Analysis of the synchronization requirements of {5G} and corresponding solutions,'' \emph{IEEE Communications Standards Magazine}, vol.~1, no.~1, pp. 52--58, 2017.

\bibitem{del2017survey}
J.~A. del Peral-Rosado \emph{et~al.}, ``Survey of cellular mobile radio localization methods: From {1G} to {5G},'' \emph{IEEE Communications Surveys \& Tutorials}, vol.~20, no.~2, pp. 1124--1148, 2017.

\bibitem{neinavaie2021acquisition}
M.~Neinavaie \emph{et~al.}, ``Acquisition, {D}oppler tracking, and positioning with starlink {LEO} satellites: First results,'' \emph{IEEE Transactions on Aerospace and Electronic Systems}, vol.~58, no.~3, pp. 2606--2610, 2021.

\bibitem{dureppagari2023ntn-arxiv}
H.~K. Dureppagari \emph{et~al.}, ``Ntn-based {6G} localization: Vision, role of {LEOs}, and open problems,'' \emph{arXiv preprint arXiv:2305.12259}, 2023.

\bibitem{ali1998doppler}
I.~Ali \emph{et~al.}, ``{D}oppler characterization for {LEO} satellites,'' \emph{IEEE transactions on communications}, vol.~46, no.~3, pp. 309--313, 1998.

\bibitem{harounabadi2023toward}
M.~Harounabadi \emph{et~al.}, ``Toward integration of {6G}-{NTN} to terrestrial mobile networks: Research and standardization aspects,'' \emph{IEEE Wireless Communications}, vol.~30, no.~6, pp. 20--26, 2023.

\bibitem{geraci2022integrating}
G.~Geraci \emph{et~al.}, ``Integrating terrestrial and non-terrestrial networks: {3D} opportunities and challenges,'' \emph{IEEE Communications Magazine}, vol.~61, no.~4, pp. 42--48, 2022.

\bibitem{charbit2021satellite}
G.~Charbit \emph{et~al.}, ``Satellite and cellular networks integration-a system overview,'' in \emph{2021 Joint European Conference on Networks and Communications \& {6G} Summit (EuCNC/{6G} Summit)}.\hskip 1em plus 0.5em minus 0.4em\relax IEEE, 2021, pp. 118--123.

\bibitem{sallouha2024ground}
H.~Sallouha \emph{et~al.}, ``On the ground and in the sky: {A} tutorial on radio localization in ground-air-space networks,'' \emph{IEEE Communications Surveys \& Tutorials}, 2024.

\bibitem{gonzalez20235g}
A.~Gonzalez-Garrido \emph{et~al.}, ``{5G} positioning reference signal configuration for integrated terrestrial/non-terrestrial network scenario,'' in \emph{2023 IEEE/ION Position, Location and Navigation Symposium (PLANS)}.\hskip 1em plus 0.5em minus 0.4em\relax IEEE, 2023, pp. 1136--1142.

\bibitem{gonzalez2024interference}
------, ``Interference analysis and modeling of positioning reference signals in {5G} {NTN},'' \emph{IEEE Open Journal of the Communications Society}, 2024.

\bibitem{liang2024toward}
T.~Liang \emph{et~al.}, ``Toward seamless localization and communication: {A} satellite-{UAV NTN} architecture,'' \emph{IEEE Network}, 2024.

\bibitem{jin2024fusion}
T.~Jin \emph{et~al.}, ``Fusion positioning of {GNSS}-cellular signals of opportunity under low observability,'' \emph{IEEE Transactions on Instrumentation and Measurement}, 2024.

\bibitem{zhang2020joint}
F.~Zhang \emph{et~al.}, ``Joint range and velocity estimation with intrapulse and intersubcarrier {D}oppler effects for {OFDM}-based radcom systems,'' \emph{IEEE Transactions on Signal Processing}, vol.~68, pp. 662--675, 2020.

\bibitem{prasad2004ofdm}
R.~Prasad, ``{OFDM} for wireless communications systems,'' \emph{Artech House Universal Personal Communications Library/Artech House}, 2004.

\bibitem{riley2008handbook}
W.~J. Riley \emph{et~al.}, ``Handbook of frequency stability analysis,'' 2008.

\bibitem{Kamran_JSTSP_SISO_RIS}
K.~Keykhosravi \emph{et~al.}, ``{RIS}-enabled {SISO} localization under user mobility and spatial-wideband effects,'' \emph{IEEE Journal of Selected Topics in Signal Processing}, vol.~16, no.~5, pp. 1125--1140, 2022.

\bibitem{etsi2020138}
T.~ETSI, ``138 211-v16. 2.0, {5G NR},'' \emph{Physical channels and modulation (3GPP TS 38.211 version 16.2. 0 Release 16), ETSI}, 2020.

\bibitem{koivunen2024multicarrier}
V.~Koivunen \emph{et~al.}, ``Multicarrier {ISAC}: Advances in waveform design, signal processing, and learning under nonidealities,'' \emph{IEEE Signal Processing Magazine}, vol.~41, no.~5, pp. 17--30, 2024.

\bibitem{alkhateeb2016frequency}
A.~Alkhateeb \emph{et~al.}, ``Frequency selective hybrid precoding for limited feedback millimeter wave systems,'' \emph{IEEE Transactions on Communications}, vol.~64, no.~5, pp. 1801--1818, 2016.

\bibitem{wymeersch2022radio}
H.~Wymeersch \emph{et~al.}, ``Radio localization and sensing—part i: Fundamentals,'' \emph{IEEE Communications Letters}, vol.~26, no.~12, pp. 2816--2820, 2022.

\bibitem{ercan2024ris}
M.~K. Ercan \emph{et~al.}, ``{RIS}-aided {NLoS} monostatic multi-target sensing under angle-{D}oppler coupling,'' \emph{Authorea Preprints}, 2024.

\bibitem{Keykhosravi2020_SisoRIS}
K.~Keykhosravi \emph{et~al.}, ``{SISO} {RIS}-enabled joint {3D} downlink localization and synchronization,'' in \emph{IEEE Int. Conf. Commun.}, 2021, pp. 1--6.

\bibitem{stutzman2012antenna}
W.~L. Stutzman \emph{et~al.}, \emph{Antenna theory and design}.\hskip 1em plus 0.5em minus 0.4em\relax John Wiley \& Sons, 2012.

\bibitem{Fortunati2017}
S.~Fortunati \emph{et~al.}, ``Performance bounds for parameter estimation under misspecified models: {F}undamental findings and applications,'' \emph{IEEE Signal Process. Mag.}, vol.~34, no.~6, pp. 142--157, 2017.

\bibitem{Ricmond2015MCRB}
C.~D. Richmond \emph{et~al.}, ``Parameter bounds on estimation accuracy under model misspecification,'' \emph{IEEE Trans. Signal Process.}, vol.~63, no.~9, pp. 2263--2278, 2015.

\bibitem{richards2022fundamentals}
M.~A. Richards \emph{et~al.}, \emph{Fundamentals of radar signal processing}.\hskip 1em plus 0.5em minus 0.4em\relax Mcgraw-hill New York, 2022, vol.~1.

\bibitem{kay1993fundamentals}
S.~M. Kay, \emph{Fundamentals of statistical signal processing: estimation theory}.\hskip 1em plus 0.5em minus 0.4em\relax Prentice-Hall, Inc., 1993.

\end{thebibliography}

\end{document}